\documentclass[conference,compsoc]{IEEEtran}
\usepackage[numbers,sort&compress]{natbib} %
\usepackage{xurl}
\usepackage{xcolor} 
\usepackage[T1]{fontenc}
\usepackage[utf8]{inputenc}
\usepackage[scaled=0.98]{inconsolata}   %
\usepackage[htt]{hyphenat}              %

\newif\ifcomment
\commenttrue 
\commentfalse
\ifcomment
	\newcommand{\edc}[1]{\textbf{\em\color{red}EDC: #1}}

	\newcommand{\gts}[1]{\textbf{\em\color{green}GTS: #1}}
	\newcommand{\chloe}[1]{\textbf{\em\color{blue}Chloe: #1}}
    \definecolor{goldenrod}{rgb}{0.85, 0.65, 0.13}
    \newcommand\christoph[1]{\textcolor{goldenrod}{-- Christoph: #1 --}}
\else
	\newcommand{\edc}[1]{}

	\newcommand{\gts}[1]{}
	\newcommand{\chloe}[1]{}
    \newcommand\christoph[1]{}
\fi
\usepackage{pifont}
\usepackage{threeparttable}
\usepackage{tabularx} 

\definecolor{darkred}{RGB}{150, 0, 0}
\definecolor{darkblue}{RGB}{0,0,110}
\usepackage[bookmarks=false, colorlinks=true, plainpages=false,  linkcolor=darkred, citecolor=darkred, urlcolor=darkblue, filecolor=blue]{hyperref}

\usepackage[hang,flushmargin]{footmisc}
\renewcommand{\footnoterule}{%
  \kern-3pt
  \hrule width 0.5\columnwidth height 0.4pt
  \kern 3pt
}

\makeatletter
\def\url@leostyle{%
  \@ifundefined{selectfont}{\def\UrlFont{}}%
  {\def\UrlFont{}}%
}
\makeatother
\usepackage{xurl}                     %
\pdftrailerid{}          %

\newcommand{\dataset}{25\xspace}

\usepackage{graphicx}
\usepackage{amsmath}
\usepackage{booktabs}
\usepackage{enumitem}
\usepackage[font=small,labelfont=bf,labelsep=period,position=below]{caption}

\usepackage{etoolbox}

\usepackage{capt-of} %
\usepackage{xspace}

\usepackage{array}
\newcolumntype{H}{>{\setbox0=\hbox\bgroup}c<{\egroup}@{}}
\newcolumntype{P}[1]{>{\raggedright\arraybackslash}p{#1}}

\newcommand\descr[1]{\smallskip\noindent\textbf{#1}}
\newcommand{\perm}[1]{{\def\_{\textunderscore\allowbreak}\textsc{#1}}}

\begin{document}

\title{\bf What's On Your Mind? Exploring Privacy of Mental Health Apps}

\author{Chloe Georgiou$^1$, Hans Lu$^1$, Emiliano De Cristofaro$^2$, Gene Tsudik$^1$ \\[0.25em]
$^1$University of California, Irvine \quad $^2$University of California, Riverside}

\maketitle
\thispagestyle{plain}

\begin{abstract}
Therapy and life-coaching apps have grown rapidly in popularity, yet their users often share highly sensitive information, including mental health issues, trauma, and relationship difficulties. This prompts the need to examine privacy practices across the ecosystem. In this paper, we present a comprehensive analysis of 25 popular Android mental health and life-coaching apps, combining static analysis, dynamic network traffic analysis, and identifying gaps between observed app behavior and privacy policies.

Our analysis highlights serious concerns and substantial transparency gaps. First, all but one app embeds at least one tracker SDK not named in its privacy policy, and 90\% of the apps we instrument fail to disclose at least half of their detected trackers. Second, more than half of the apps declare dangerous permissions without corresponding privacy-policy disclosures, including camera or microphone permissions. Third, 14 of the 25 apps disclose third-party AI processing, but only five name the providers (e.g., OpenAI, Anthropic, and Groq), while the others use only generic language (e.g., ``AI services'').

Overall, our results highlight shortcomings in current disclosure practices that undermine meaningful informed consent. We argue for stronger privacy protections for therapy apps, more closely aligned with the confidentiality expectations associated with licensed human therapists.
\end{abstract}
\IEEEpeerreviewmaketitle 

\sloppy

\section{Introduction}
The digital mental health ecosystem has evolved from niche wellness tools to a more mainstream app presence~\cite{torous2025evolving}, with a market projected to grow from the current \$8.2B to \$40.9B by 2035~\cite{gminsights2026mentalhealthapps}.
Popular examples include Headspace, a meditation and mindfulness app with millions of paid subscribers~\cite{businessofapps2025headspace}, Replika, an AI-companion app with a user base reported at 40 million in 2025~\cite{businessinsider2025replika}, and 7~Cups, an app claiming to have connected over 70 million users with peer supporters and licensed clinicians~\cite{7Cups2025stats}.

Mental health tools, often coupled with generative AI and/or wearables, can enable self-help, coach-guided, or clinician-led interventions beyond the constraints of traditional telehealth~\cite{torous2025evolving}.
Therapy apps promise to reduce structural barriers to conventional psychotherapy, such as cost and limited insurance coverage~\cite{starvaggi2025psychotherapy} as well as off-hours availability~\cite{baumel2018digital,coffield2025evaluating}.
Moreover, unlike human practitioners, who may be subject to mood fluctuations, fatigue, and biases that can impair clinical judgment and therapeutic alliance~\cite{sayer2024clinician}, therapy apps promise consistent therapeutic feedback and engagement.

\descr{Motivation.} Therapy apps occupy a uniquely sensitive operational context, since users may routinely disclose highly intimate details, e.g., traumas, suicidal thoughts, interpersonal conflicts, and clinical diagnoses.
Many users would not disclose such sensitive information even to family members or close friends.
Therapists are bound by professional codes and typically hold professional credentials that attest to their formal training and competency examinations~\cite{apa2017ethics,aca2014ethics}.
Meanwhile, therapy apps are not bound by the same rules or standards~\cite{torous2017needed,fowler2022mind}, and in the U.S., most fall outside HIPAA, which regulates how licensed practices handle and transfer patient data~\cite{kim2023databrokers}.
Unauthorized disclosure of patient information constitutes a serious ethical and legal violation for licensed therapists, whereas therapy apps are not inherently prevented from transmitting user (meta-)data to third-party entities, e.g., advertising networks, AI providers, and social media platforms.

Unsurprisingly, therapy apps increasingly include AI functions, e.g., chatbot interactions.
Sensitive personal disclosures accumulated across longitudinal sessions may be retained on users' devices and/or stored in cloud-based infrastructures, a practice not fundamentally distinct from electronic health records ubiquitous among practitioners.
However, user devices are more vulnerable to security threats, including unauthorized access to, leakage of, or modification of therapy records.
One prominent (and terrifying) example of the dire consequences of mental health data leakage is the 2020 data breach of the Finnish psychotherapy provider Vastaamo~\cite{wikipedia2024vastaamo}.
After cybercriminals copied the entire patient database and demanded a ransom, the provider refused, and records of 36k patients were released.
Many of these patients were also extorted, extremely traumatized, and a few even committed suicide~\cite{looi2024vastaamo}.

\descr{Roadmap.} These concerns prompt the need to analyze privacy practices of the digital mental health and life-coaching app ecosystem.
More precisely, we need to identify the third-party services involved in these apps (e.g., analytics, advertisers, AI providers) and compare each app's actual behavior with its privacy policy disclosures.
To tackle these issues, this paper presents a multi-method measurement study of 25 popular Android apps offering mental health and life-coaching services.
The study combines static, dynamic, and semantic privacy policy analyses to identify security and privacy issues and investigate whether and how apps' policies accurately disclose their observable data practices.

On the behavioral side, we use two complementary static analysis tools, MobSF~\cite{mobsf} and Androguard~\cite{androguard}, to extract dangerous permissions declared in Android manifests, identify embedded third-party tracker SDKs, and assess security configurations.
We also capture and analyze network traffic with mitmproxy~\cite{mitmproxy} and PCAPdroid~\cite{pcapdroid} during instrumented application walkthroughs, uncovering third-party domains contacted at runtime.
On the policy side, we extract structured privacy practice claims through a two-stage pipeline: an initial keyword-matching pass followed by LLM-assisted extraction using Gemini 3.1 Pro, validated against manual annotation and cross-checked with GPT-5.4.

We compare apps' behavior with policy claims across: (1) disclosure of embedded trackers, (2) dangerous permissions, and (3) third-party AI providers.
Overall, the goal of our measurements is to answer the following questions:
\begin{enumerate}[leftmargin=*, topsep=2pt, itemsep=1pt, parsep=1pt]  
  \item %
To what extent are the observed data practices of popular therapy apps accurately disclosed in their corresponding privacy policies?
  \item %
How does any disclosure gap differ across embedded third-party trackers, dangerous permissions, and third-party AI providers?
  \item %
Are the disclosure gaps we measure robust to errors in automated policy extraction?
\end{enumerate}

\descr{Key Findings.} Our results raise several concerns:
\begin{itemize}[leftmargin=*, topsep=2pt, itemsep=1pt, parsep=1pt]  
\item \textit{Significant disclosure gap:} Static analysis with MobSF detects in 24 out of 25 apps {\it at least} one SDK-embedded tracker that is not named in the app's privacy policy.
Also, 11/21 apps that request camera or microphone permissions fail to disclose at least one of those permissions in their privacy policy.

\item \textit{Dynamic analysis reveals runtime behavior invisible to static analysis:} Since 5/25 apps require organizational access codes, we could not exercise their full functionality.
Of the other 20 apps, which we instrumented at runtime, 19 contact at least one third-party tracker domain not named in their policy.
For example, {\it Talkie} contacts 8 distinct advertising domains (AdMob, AppLovin, BidMachine, Fyber, InMobi, Mintegral, TikTok/Pangle, Unity), while disclosing none.

\item \textit{Vast majority of apps fail to disclose trackers:} 18/20 fail to disclose at least half of embedded trackers, e.g., {\it Talkie} embeds 21 trackers while its policy discloses none.
\item \textit{Many apps use third-party AI:} 14/25 apps' privacy policies state that user data is processed by third-party AI providers.
Only 5 name specific providers; e.g., Rosebud names OpenAI, Anthropic, and Groq.
The remaining 9 use generic language, such as ``AI Services'' or ``Large Language Models.''
\end{itemize}

\descr{Implications.}
While concerning, our findings {\bf should not be interpreted} as direct evidence of deliberate misconduct by app developers.
Identified issues may be attributable to common systemic factors, e.g., inheriting third-party SDKs without adequate privacy auditing~\cite{razaghpanah2018apps,reardon2019fifty,kollnig2021consent, alomar2025effect}, reusing boilerplate policy language~\cite{andow2019policylint}, or developer unawareness of GDPR's consent obligations~\cite{nguyen2021share}.

Nonetheless, our study raises serious concerns for an inherently vulnerable population.
Therapy app users routinely disclose their intimate thoughts and feelings while expecting privacy (i.e., strict confidentiality, as with human therapists) and security (e.g., integrity)~\cite{torous2017needed,kwesi2025exploring} that apps are not mandated to provide~\cite{torous2017needed,huckvale2019assessment}.
Addressing this gap requires structural interventions, e.g., mandatory disclosure of trackers, standardized privacy auditing requirements, and enforceable data minimization obligations.
Our work also suggests that disclosures must be mandatory to be meaningful, and we hope it will provide empirical grounding for adequate regulatory developments.

\descr{Paper Organization.}
Section~\ref{sec:dat} describes our dataset of 25 therapy and wellness apps.
Then, Section~\ref{sec:methodology} presents our methodology, %
while Section~\ref{sec:results} reports our results on observed app behavior, the claims made in privacy policies, and the resulting disclosure gaps.
Section~\ref{sec:rel} reviews related work, while Section~\ref{sec:discussion} discusses the %
implications of our findings.
Section~\ref{sec:conclusion} concludes the paper.

\section{Dataset}\label{sec:dat}
As is common in prior work~\cite{gorla2024analysis, reardon2019fifty, kollnig2021iphones}, we focus on the Android platform for two main reasons: (1) its dominance of the global mobile OS market (72\% share as of November 2026~\cite{statcounter2025}), and (2) its open APK format, which facilitates static analysis.

\descr{Systematic App Discovery.} 
We identify candidate apps through Google Play searches on the US store, in English, using the following keywords: %
{\em therapy, mental health, CBT, counseling, emotional wellness.}\footnote{CBT stands for Cognitive Behavioral Therapy.}
These terms span both user-facing language (how people search for support) and clinically grounded terminology (established intervention types); prior app studies select candidates through similar keyword searches~\cite{alqahtani2020insights,marshall2020apps,haque2022app}.
(We did not record the number of results per keyword.)
We retrieve the number of Google Play downloads and retain apps with at least $10$k downloads, following prior privacy studies of Android apps that focus on widely installed apps~\cite{nguyen2021share,reyes2018coppa, reardon2019fifty}.
This threshold excludes apps with negligible user bases while keeping the corpus small enough to exercise each app manually and annotate each policy by hand.
Incidentally, prior work on mental health apps finds that Google Play's install bands correlate strongly with independently measured download counts (Spearman $r=.81$, $N=93$, $p<.001$), supporting their use as a popularity measure~\cite{baumel2019objective}.

Our English-language keyword search returns no apps that only support other languages and exceed $10$k downloads.
Finding such apps would require keywords in languages we are not fluent in.
While several Chinese-language apps for mental health support~\cite{restofworld2025aitherapychina,yin2020china} are reportedly available from Chinese app stores, e.g., KnowYourself, Jiandanxinli, and JD~Health's AI therapeutic companion, we leave a multilingual analysis of therapy app privacy practices to future work.

\begin{figure}[t]
 \centering
 \includegraphics[width=0.9999\columnwidth]{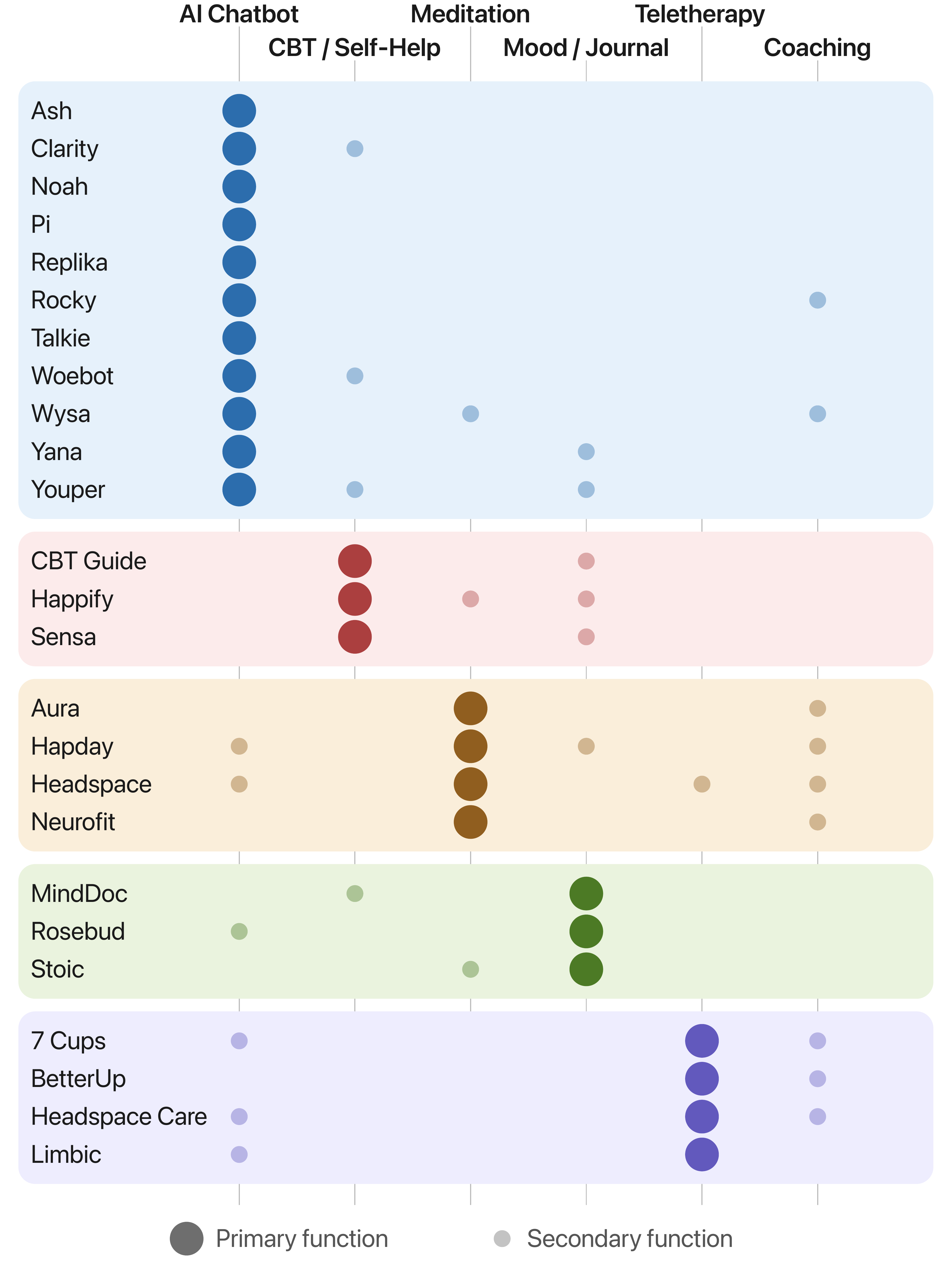} 
 \caption{Primary and secondary functions of 25 apps in our corpus, denoted by dark/big and light/small circles.}
 \label{fig:app-categories}
 \end{figure}

We also exclude two apps that meet our criteria but are no longer available: Sintelly (1M+ downloads) and ChatMind (5M+).
The final app dataset, summarized in Figure~\ref{fig:app-categories}, spans five functional categories:
\begin{itemize} %
\item \emph{AI Companions and Chatbots (10):} Replika, Talkie, Pi, Rocky, Ash, Noah, Woebot, Wysa, Youper, Yana;
\item \emph{CBT and Self-Help (4):} Clarity, Happify, Sensa, CBT Guide to Depression \& Test (hereafter, CBT Guide);
\item \emph{Meditation/Wellness (4):} Headspace, Aura, Neurofit, Hapday;
\item \emph{Journaling and Mood Tracking (3):} Stoic, Rosebud, MindDoc;
\item \emph{Teletherapy and Coaching (4):} 7~Cups, Headspace Care (formerly known as Ginger), BetterUp, Limbic. 
\end{itemize}
Table~\ref{tab:app-overview} in Appendix~\ref{appendix:app-details} provides a more detailed overview of the 25 apps used in this study, including a short description of their functionalities, number of downloads, year of first appearance on Google Play, developer's country, etc.

\descr{App Corpus.} 
The apps in the dataset span AI companions, self-help tools, and teletherapy.
Install counts span three orders of magnitude, including small apps (such as Neurofit, Hapday, and Limbic with 10K+ installs) to mass-market platforms such as Headspace, Replika, Talkie, and Yana, with 10M+ downloads.

The way users interact with these apps also varies widely.
Some primarily offer a user interface to AI chatbots: Replika, Talkie, Pi, Ash, Noah, Rocky, Yana, Youper, Clarity, Rosebud, Neurofit, Hapday.
Meanwhile, others provide self-guided libraries of lessons, journals, or meditations with no real-time chatbot and no set time frame for completion: Happify, Sensa, CBT~Guide, Stoic, MindDoc, and Aura.
Another group blends an AI chatbot with optional access to human coaches or licensed therapists: Wysa, Headspace, BetterUp, 7~Cups, and Headspace Care.
Some others (Woebot and Limbic) use a chatbot that feeds into a human clinician's workflow.
Pricing is similarly all over the map: while Pi and Ash are free, most apps follow a {\it freemium model}.
Sensa and Neurofit are subscription-only, and BetterUp and Headspace Care are typically paid by employers or insurers.

\descr{Access to App Functionality.} 
Five apps have restricted access: Woebot and Limbic require invitation codes from a healthcare provider and an organization, respectively, while Headspace, Headspace Care, and BetterUp require an organizational account.
This limits their analysis to pre-authentication functionality.
However, we retain them in the corpus because their APKs and privacy policies can still be analyzed.
Although tracker counts for these five apps may underestimate their full runtime behavior, we report their analysis results alongside all other apps.

\descr{Data Collection.} 
In July 2026, for each app, we collect:
\begin{itemize}[leftmargin=*, topsep=2pt, itemsep=1pt, parsep=1pt]  
\item Metadata, including: download count, country of developer origin, year of first release, and the last update date from the Google Play listing.
\item The APK binary: for 22/25 apps, we downloaded it directly from Google Play using apkeep~\cite{apkeep}, authenticated with an OAuth token from a dedicated account, under a pinned device profile, locale, and timezone.
We verify each download's signing certificate with apksigner and record its version code, version name, SDK levels, SHA-256 hash, and signer certificate hash.
We run all analyses on the builds installed on our test device on July 28, 2026; Google Play had updated 7 of them to a newer version since the download.
For three apps unavailable through this method (Limbic, Youper, and Happify), we cross-reference public repositories (APKPure, APKMirror, APKCombo, and AndroZoo~\cite{androzoo}) and retain the most recent available version of each app as of July 2026.
\item The app's current privacy policy, captured as HTML from the URL listed in the Google Play Store.
\item Initial data collection: after installing each app, we attempted to create an account, and document Personally Identifiable Information (PII) it requests before account creation or first use, authentication options it offers, and any explicit privacy assurances, e.g., ``your data is private.''
\end{itemize}
We successfully obtain all privacy policies; no app is excluded due to missing or unreachable policies.
Also, APK and privacy policy versions match for all apps.

\section{Methodology}
\label{sec:methodology}
\begin{figure}[t]
    \centering
    \includegraphics[width=0.9999\columnwidth]{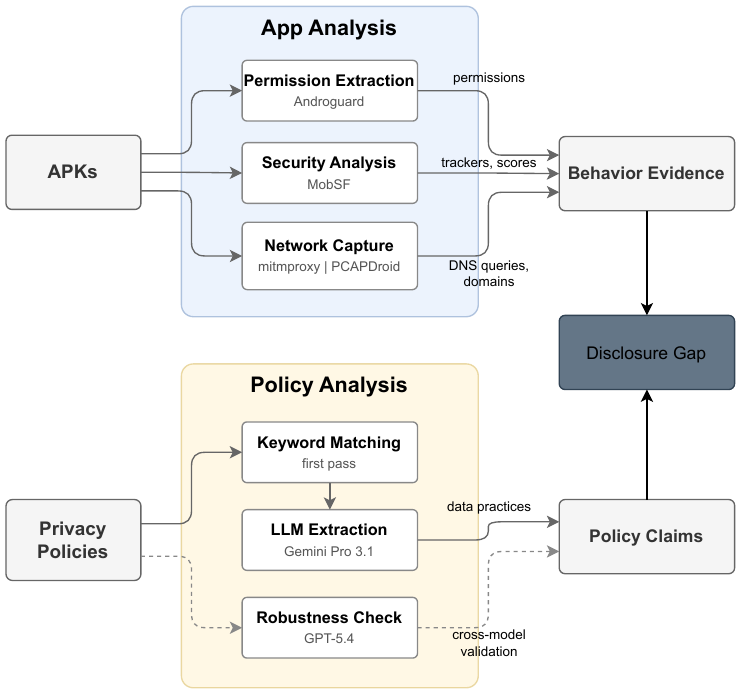}
    \caption{Overview of the measurement pipeline: APKs are analyzed statically and dynamically to extract app behavior, which is then compared with structured policy claims extracted from apps' privacy policies.}
\label{fig:pipeline}
\end{figure}

We now describe our measurement pipeline, summarized in Figure~\ref{fig:pipeline}.
It includes static and dynamic analysis to extract app security/privacy behavior, LLM-supported extraction of privacy policy claims, and an analysis of disclosure gaps.

\subsection{Static Analysis}\label{subsec:static}
We use two open-source static analysis tools: Androguard~\cite{androguard} and MobSF~\cite{mobsf} (v4.4.3).
We ran Androguard on July 28, 2026, and MobSF -- on July 31, 2026 (August 22 for Limbic).
Both are widely used in prior Android privacy and security measurements ~\cite{zimmeck2017automated,wessels2025hytrack,279902,kumar2020security,kishnani2023assessing}.
These two tools serve different purposes: Androguard extracts permissions, while MobSF covers trackers, cryptographic checks, and other security checks in a single pipeline.

\descr{Permission Extraction.}
We use Androguard to parse each app's \texttt{AndroidManifest.xml} and extract all declared dangerous permissions.
Our primary disclosure analysis centers on \perm{camera} and \perm{record\_audio} since they are most directly tied to sensitive data capture in the therapy context, e.g., recording sessions, voice journaling, and video calls.

\descr{Declared vs.~Exercised Permissions.} 
We treat a declared dangerous permission as a disclosable data category regardless of whether the app exercises it.
We do so because GDPR's Article~13(1)(c) requires informing users, ``at the time when personal data are obtained,'' of the ``purposes of the processing for which the personal data are intended''~\cite{gdpr}.
Thus, this duty attaches to intended processing, not only completed processing.
Ultimately, a declared \perm{camera} or \perm{record\_audio} permission represents a capability which the user is asked to grant.
Although this reading is partly normative, we consider privacy policies as {\em ex-ante notices}, i.e., statements made before collection about what an app may do.

\descr{Security Analysis.}
We use MobSF for automated static analysis of Android APKs, checking for manifest misconfigurations, insecure cryptographic implementations, hard-coded secrets, and known vulnerability patterns.
MobSF tags findings with Common Weakness Enumeration (CWE) identifiers and severity labels (high, warning, or good)~\cite{mobsf-rules}.
We use MobSF primarily to identify trackers and analyze manifests.
Its composite App Security Score is a weighted count of rule hits rather than a validated security metric; thus, we will report these scores as descriptive context only.

\subsection{Dynamic Traffic Analysis}
For each app, we captured its network traffic at runtime on a Samsung Galaxy~S24, between July 31 and August 5, 2026.
Because many apps pin TLS certificates, we combine active interception with passive on-device capture.
If an app allows it, we intercept traffic actively, routing it through mitmproxy~\cite{mitmproxy}.

\descr{Pinning.} Since many apps use certificate pinning (i.e., they are configured to accept only a specific TLS certificate, preventing any intercepting proxy from decrypting the connection), we use apk-mitm~\cite{apkmitm} to circumvent pinning.
This tool repackages and re-signs the APK to trust our certificate authority, allowing us to decrypt the full request and observe which server an app contacts as well as what it sends.
We tag user-entered text with a canary phrase (\texttt{PURPLE-CANARY-1234}) to trace where it travels.

If certificate pinning is implemented within the application, for example, using the OkHttp library's \texttt{CertificatePinner} class, repackaging alone may not remove the check.
For these apps, we attempt to bypass the check at runtime using Frida~\cite{frida} and the Frida-based objection toolkit~\cite{objection}.
Out of $20$ apps for which we could capture network traffic (recall that $5$ are gated by an invitation code), we obtain decrypted application-layer traffic for $8$ by bypassing TLS protections, including certificate pinning (if present), with apk-mitm.

However, 12 apps resist active interception.
For these, we capture traffic passively via PCAPdroid~\cite{pcapdroid}, an on-device local VPN that records every connection, attributes it to the responsible app through its socket UID, and recovers the destination from the cleartext TLS SNI and DNS query.
Though passive capture cannot read payloads, it leaves the app unmodified and its anti-tampering defenses never activate.
This is essential for Wysa, which uses the MaxGuard library~\cite{maxguard} to detect Frida and rooting, and halt execution.

\descr{Walkthrough.} 
For each app, we manually exercise its primary features while capturing traffic.
Depending on the app, this includes: signing in, starting a conversation or session, completing a mood check-in or journal entry, and browsing in-app content.
We enter the canary phrase into the 11 apps that offer an AI chatbot, as well as 7~Cups.
In the 5 apps we could both exercise and decrypt (7~Cups, Noah, Rocky, Rosebud, and Talkie), the canary appears in traffic to the app's own servers.
For apps whose traffic is captured passively, the encrypted payload is not observable.

\descr{Limitations.} 
We capture network traffic to identify third-party domains contacted at runtime, corroborating and supplementing the tracker measurements from static analysis.
We report which third parties each app contacts, not what is sent; while we use decrypted payloads to trace the canary phrase and identify AI and cloud providers, we do not learn the identifiers each tracker receives and we cannot observe payloads for the 12 apps captured passively.
Naturally, we can only observe the first hop from the device, e.g., we can identify an AI provider when an app calls it directly or names it in a client-side payload, while connections routed through an app's backend remain invisible.

Despite capturing network traffic for all 25 apps, Woebot, Headspace, Headspace Care, BetterUp, and Limbic require organizational access.
Thus, we can only capture their activity before the user authentication phase.
Consequently, tracker counts for these apps (based on dynamic analysis) likely underestimate their actual runtime behavior.
For completeness, we report traffic captures for these apps, though we primarily rely on static analysis.

\subsection{Privacy Policy Analysis}
\label{sec:policy-method}
Next, we extract structured privacy claims from each app's privacy policy.

\descr{Keyword Matching.} 
As a first pass, we run each policy through domain-specific lexicons covering five categories: (1)~camera or photo data collection, (2)~microphone or audio data collection, (3)~AI/LLM service providers, (4)~named trackers and analytics services, and (5)~advertising and marketing data sharing.
For category (3), we use two disjoint keyword sets (case-insensitive):
\begin{enumerate}[leftmargin=*, topsep=2pt, itemsep=1pt, parsep=1pt]  
\item \emph{Named Providers:} e.g., ``openai,'' ``chatgpt,'' ``gpt,'' ``anthropic,'' ``claude,'' ``groq.''
\item \emph{Generic AI Terminology:} e.g., ``large language model,'' ``llm,'' ``artificial intelligence,'' ``machine learning,'' ``ai service provider.''
\end{enumerate}
This distinction is methodologically important because GDPR Article~13(1)(e) requires disclosing the recipients/categories of recipients of personal data~\cite{gdpr,wp29}: a policy that names a specific provider identifies an external data recipient whose handling practices users can independently review.
In contrast, generic descriptions (e.g., ``we may use artificial intelligence to improve our services'') leave the third party unidentified.

\descr{LLM-Assisted Extraction.} 
Because privacy-policy claims are typically buried in dense and hard-to-comprehend legal language, keyword matching alone misses indirect disclosures.
Therefore, inspired by Xie et al.~\cite{xie2025evaluating}, we run each policy through two LLMs: Gemini 3.1 Pro (\texttt{gemini-3.1-pro-preview}) and GPT-5.4 (\texttt{gpt-5.4-2026-03-05}).
We provide each policy as an HTML attachment along with a classification prompt (see Appendix~\ref{appendix:prompt}) that asks the model to return, for each of the five categories above, a binary label and supporting verbatim quotes from the policy text.

\descr{Ground Truth and Validation.} 
To validate this approach, two authors independently annotate 24 unique privacy policies in our corpus (Headspace and Headspace Care share one) across the same five categories, agreeing on 110 of the 120 labels (Cohen's $\kappa=0.83$).
We resolve the ten disagreements by re-examining the policy text, correcting one label that both initially missed, and running both models against the reconciled labels.
We report the results in Section~\ref{sec:rq3}.

\section{Results}
\label{sec:results}
We now present study results, focusing on three outputs of the measurement pipeline, as shown in Figure~\ref{fig:pipeline}: (1) behavior evidence from app analysis, (2) policy claims, and (3) the disclosure gap found by comparing them.
 
\subsection{Behavior Evidence}
\label{sec:behavior-evidence}
We present results of static and dynamic analyses across 4 dimensions: (1) PII and privacy assurances recorded before account creation or first use; (2) dangerous permissions Androguard extracts from each APK's \texttt{AndroidManifest.xml}; (3) embedded third-party tracker SDKs flagged by MobSF; and (4) third-party domains each app contacts at runtime, based on mitmproxy and PCAPdroid traffic capture.

Static analysis (permissions and embedded tracker SDKs) covers all 25 apps, while runtime traffic capture requires exercising the app and only covers 20 apps usable without organizational access codes.
In Table~\ref{tab:coverage}, we summarize which analysis is applied to which set of apps.

\begin{table}[t]
\centering
\small
\setlength{\tabcolsep}{6pt}
\begin{tabular}{lrl}
\toprule
\textbf{Analysis} & $n$ & \textbf{Excluded} \\
\midrule
Static Analysis & 25 & -- \\
Dynamic Analysis          & 20 & 5 gated by org.\ code \\
PII at first use                   & 20 & 5 gated by org.\ code \\
Camera/Mic Disclosure & 21 & 4 declare neither\\
\bottomrule
\end{tabular}
\caption{Corpus coverage per analysis.}
\label{tab:coverage}
\end{table}

\descr{Initial Data Collection.} %
\label{sec:initial-collection-results}
After installing each app, we document all PII requested before account creation or first use.
We do so since these apps may ask for sensitive data, such as mental-health screening answers, age (and/or date-of-birth), or gender, i.e., requiring users to share PII information up front, before they can fully understand the app's functionalities and decide whether they would like to actually use it by creating an account.
We record whether each app requests an email address, name, age/date-of-birth, gender, and mental health screening questions, as well as the available authentication methods.
We also record any explicit privacy assurances when using the app, e.g., ``your data is private.''

\begin{table}[!t] 
\centering
\footnotesize 
\setlength{\tabcolsep}{-1pt}
\begin{threeparttable}
\begin{tabular}{@{}lcccccl@{}}
\toprule
\rotatebox{45}{\textbf{App}} & \rotatebox{45}{\textbf{Email}} & \rotatebox{45}{\textbf{Name}} & \rotatebox{45}{\textbf{Age/DOB}} & \rotatebox{45}{\textbf{Gender}}\hspace*{-0.1cm} & \rotatebox{45}{\textbf{MHQs}}\hspace*{-0.1cm} & \rotatebox{45}{\textbf{Auth}} \\
\midrule
\multicolumn{7}{l}{\textit{Fully Accessible Apps (20)}} \\
\midrule
Wysa               & \ding{51} & Nick      & Range      & \ding{55}         & \ding{51} & Email \\
Youper             & \ding{51} & \ding{55}            & Range      & \ding{55}         & \ding{51} & Google/Apple/Email \\
Replika            & \ding{51} & \ding{55}            & \ding{55}         & \ding{55}         & \ding{55}         & Google/Apple/Email \\
Talkie             & \ding{55}         & \ding{55}            & Mo/Yr      & \ding{51} & \ding{55}         & Auto-created \\
7~Cups             & \ding{51} & Usern.      & \ding{55}         & \ding{55}         & \ding{55}         & Email \\
Pi AI              & \ding{51} & Pref.     & 18+        & \ding{55}         & \ding{55}         & Google/Apple \\
Rocky AI           & \ding{55}         & Nick      & \ding{55}         & \ding{55}         & \ding{55}         & Google/Apple/FB/Anon \\
Ash AI             & \ding{51} & \ding{55}            & \ding{55}         & \ding{55}         & \ding{55}         & Google/Apple \\
MindDoc            & \ding{51} & \ding{55}            & \ding{55}         & \ding{55}         & \ding{55}         & Email \\
Happify            & \ding{51} & Usern.      & Range      & \ding{51} & \ding{51} & FB/Apple/Email \\
Aura Health        & \ding{51} & First    & Range      & \ding{51} & \ding{51} & Google/Apple/Email \\
Sensa              & \ding{51} & \ding{55}            & Range      & \ding{51} & \ding{51} & Email \\
Stoic              & \ding{51} & \ding{55}            & Range      & \ding{55}         & \ding{51} & Apple ID \\
Yana               & \ding{51} & \ding{55}            & \ding{55}         & \ding{55}         & \ding{51} & Google/FB/Apple/Email \\
Rosebud            & \ding{51} & First    & \ding{55}         & \ding{55}         & \ding{55}         & Google/Apple/Email \\
Noah AI            & \ding{55}         & \ding{55}            & \ding{55}         & \ding{55}         & \ding{55}         & Email/Phone \\
Hapday             & \ding{51} & First    & \ding{55}         & \ding{55}         & \ding{51} & Apple Health required \\
Neurofit           & \ding{51} & First    & \ding{55}         & \ding{55}         & \ding{55}         & Apple/Email \\
Clarity            & \ding{51} & \ding{55}            & \ding{55}         & \ding{55}         & \ding{51} & Apple (subscription req.) \\
CBT Guide          & \ding{55}         & \ding{55}            & \ding{55}         & \ding{55}         & \ding{55}         & None (no account) \\
\midrule
\multicolumn{7}{l}{\textit{Partly or Not Accessible Apps (5)}} \\
\midrule
Headspace               & \ding{51} & Full    & \ding{55}  & \ding{55} & \ding{55} & Google/Apple/FB/Email \\
BetterUp                & \ding{51} & If Dep. & \ding{55}  & \ding{55} & \ding{55} & Google/Apple/Email/Org \\
Head.~Care\hspace{-0.1cm} & \ding{51} & Full    & DOB & \ding{55} & \ding{55} & Organization code \\
Limbic                  & \ding{51} & \ding{55}      & \ding{55}  & \ding{55} & \ding{55} & Healthcare provider only \\
Woebot & \multicolumn{6}{c}{Blocked at first screen; no PII observed} \\
\bottomrule
\end{tabular}
  \begin{tablenotes}[flushleft]
\footnotesize
    \item \hspace{-0.05cm}{\em Abbreviations:} MHQs = Mental Health Questionnaires, Nick = Nickname, Usern. = Username, Pref. = Preferred, If Dep. = If Dependent, Mo/Yr = Month/Year, FB = Facebook.
    \item[]\vspace{-5pt}\hspace*{-\labelsep}\rule{\linewidth}{0.7pt}
  \end{tablenotes}
  \end{threeparttable}
\caption{PII requested during initial use across apps.}
\label{tab:pii}
\end{table}

Table~\ref{tab:pii} reports results for 20 fully accessible apps\footnote{Recall that this excludes 5 that require an organizational code, as marked with \textsuperscript{\textdagger} in Table~\ref{tab:pii}.}, showing that 80\% (16/20) require an email address, while nearly half (9/20) ask mental health screening questions before account creation, collecting sensitive data {\bf before} a prospective user can review how it will be stored and/or shared.
We also find that 15\% of apps (3/20) request gender, age, and mental health data to use the app fully, constructing a detailed personal profile before users start using the app.
Another 20\% (4/20) explicitly reassure users that their data is private, e.g., Wysa: nicknames are ``private,'' Youper: conversations are ``private and safe,'' Happify: answers ``remain completely confidential,'' and Aura: answers are ``private and will not be shared.''
As we show later, many of these claims are contradicted by apps' actual data practices.

\descr{Permissions (Androguard).}
\label{sec:perm-observed}
Androguard extracts declared dangerous permissions from \texttt{AndroidManifest.xml}.
21/25 apps (84\%) declare \perm{camera}, \perm{record\_audio}, or both, while remaining 4 (16\%) declare neither.
These declarations are the permission baseline which we later compare against policy disclosures in Section~\ref{sec:disclosure-gap}).

\begin{figure}[t]
    \centering
    \includegraphics[width=0.9999\linewidth]{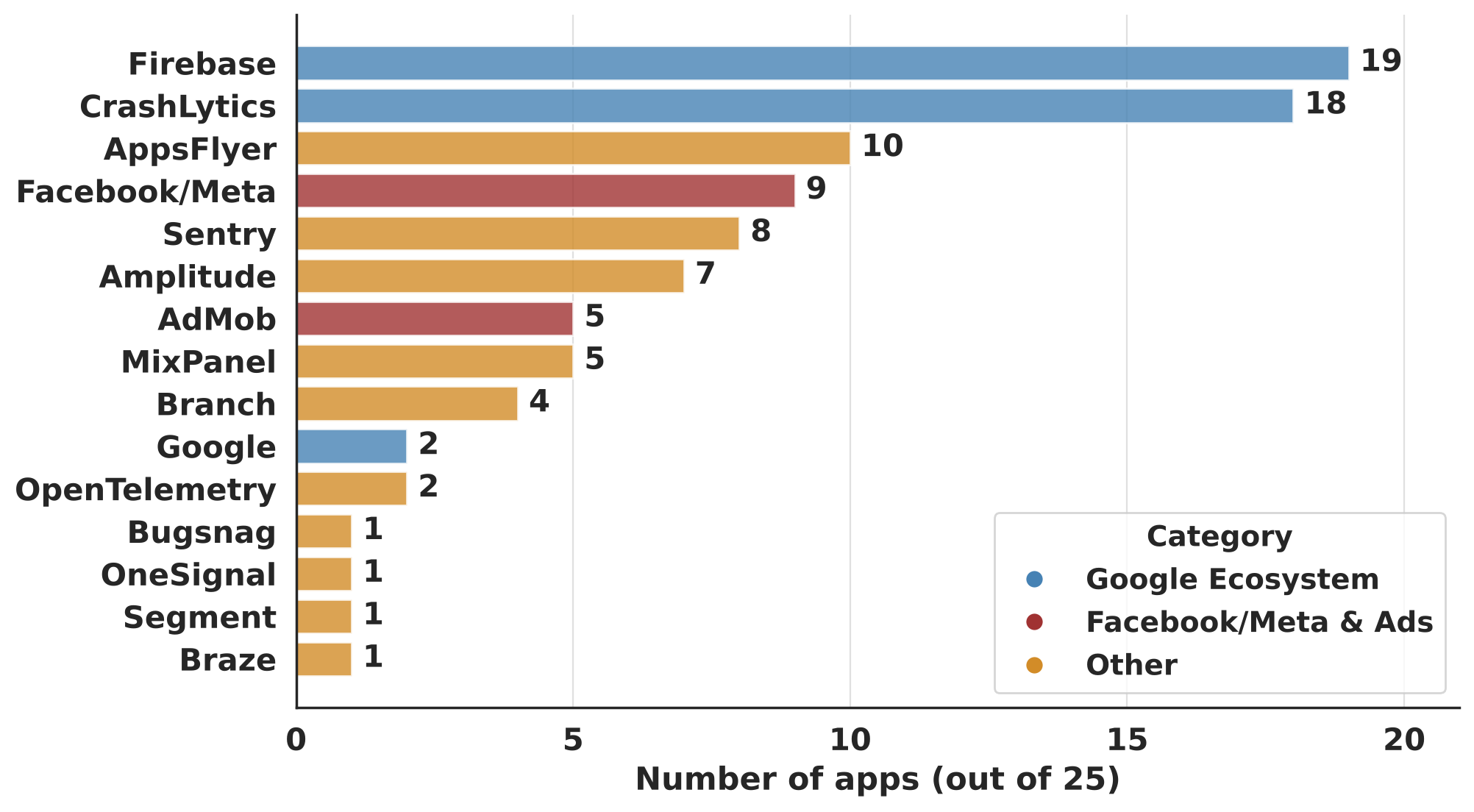} 
    \caption{Trackers detected by MobSF in 25 apps.}
    \label{fig:tracker-frequency}
\end{figure}

Besides microphone and camera, 20/25 (80\%) apps declare legacy external-storage permissions (\perm{read\_external\_storage} or \perm{write\_external\_storage}), which grant access to the photo library on Android versions 13 and earlier.
Notably, Headspace Care (a video therapy platform that connects users with licensed psychiatrists) requests photo library access.
It is unclear why it does this, since BetterUp (a comparable video coaching platform) does not, suggesting that photo library access is not fundamental to supporting video sessions.
Talkie and Replika (two popular AI-companion apps) also request full external storage access even though their primary interfaces are text/voice chat.

As mentioned in Section~\ref{subsec:static}, although declaring a permission does not mean that the app actually uses it, we believe a privacy policy must disclose what the app can access, not just what it does as a result.

\descr{Embedded Tracker SDKs (MobSF).}
\label{sec:tracker-observed}
Beyond permissions, MobSF~\cite{mobsf} identifies third-party tracker SDKs embedded in APKs by matching compiled package names against the Exodus Privacy tracker signatures: 432 trackers, list dated October 2025.
Every app embeds at least one tracker SDK.
Across 25 APKs, MobSF flags 127 SDK instances drawn from 36 distinct trackers.
The most common SDKs are: Google Firebase (19 apps), Google Crashlytics (18), AppsFlyer (10), Facebook Login (9), Facebook Share (9), Sentry (8), and Amplitude (7 apps).
(See also Figure~\ref{fig:tracker-frequency}.)
MobSF also computes a composite App Security Score per app, which we present in Appendix~\ref{app:mobsf-scores} as descriptive context rather than as a finding.

\begin{table}[t]
\centering
\setlength{\tabcolsep}{6pt}
\renewcommand{\arraystretch}{1.1} %
\begin{tabular}{@{}l >{\raggedright\arraybackslash\hyphenpenalty=10000 \exhyphenpenalty=10000}p{6.2cm}@{}}
\toprule
\textbf{App} & \textbf{Tracker Domains Observed} \\
\midrule
Talkie    & AdMob, Adjust, AppLovin, AppsFlyer, BidMachine, Crashlytics, Facebook, Firebase, Fyber, InMobi, Mintegral, Pangle, Unity \\
Aura      & AdMob, AppsFlyer, Branch, Bugsnag, Facebook, Firebase, Mixpanel, OneSignal \\
Ash       & AppsFlyer, Crashlytics, Firebase, Sentry \\
Clarity   & Amplitude, Crashlytics, Facebook, Firebase \\
Hapday    & Amplitude, AppsFlyer, Facebook, Firebase \\
Pi        & AppsFlyer, Crashlytics, Firebase, Sentry \\
Replika   & AppsFlyer, Crashlytics, Facebook, Firebase \\
Sensa     & AppsFlyer, Facebook, Firebase, Instabug \\
7~Cups    & Crashlytics, Firebase, Sentry \\
Yana      & AdMob, Crashlytics, Facebook, Firebase \\
Happify   & AppsFlyer, Crashlytics, Facebook, Mixpanel \\
Wysa      & Branch, Crashlytics, Firebase \\
Rosebud   & AppsFlyer, Mixpanel, Sentry \\
NeuroFit  & Crashlytics, Firebase \\
CBT Guide & AdMob, Crashlytics, Firebase \\
MindDoc   & Crashlytics, Firebase \\
Noah      & Amplitude, Sentry \\
Rocky     & Facebook, Firebase \\
Stoic     & Amplitude, Sentry \\
Youper    & Firebase, Mixpanel \\
\bottomrule
\end{tabular}
\caption{Dynamic analysis: embedded tracker SDKs (detected by static analysis) that we also observed contacted at runtime, from mitmproxy/PCAPdroid captures ($n=20$ apps; 5 required access codes).}
\label{tab:wireshark}
\end{table}

\begin{table}[t]
\centering
\setlength{\tabcolsep}{3pt}
\begin{tabularx}{\columnwidth}{@{}l>{\raggedright\arraybackslash}X@{}}
\toprule
\textbf{Tracker} & \textbf{Description} \\
\midrule
\multicolumn{2}{@{}l}{\textit{Analytics \& Product Insights (3)}} \\
\midrule
Firebase      & Google's mobile/web analytics and app dev platform \\
Mixpanel      & Event-based product analytics focused on user behavior funnels \\
Amplitude     & Product analytics platform tracking user journeys and retention \\
\midrule
\multicolumn{2}{@{}l}{\textit{Attribution \& Marketing (3)}} \\
\midrule
AppsFlyer     & Mobile attribution and marketing analytics platform \\
Adjust        & Mobile attribution and campaign measurement platform \\
Branch        & Deep-linking and mobile attribution platform \\
\midrule
\multicolumn{2}{@{}l}{\textit{Crash Reporting \& Monitoring (4)}} \\
\midrule
Sentry        & Real-time error tracking and crash reporting for apps \\
Crashlytics   & Firebase-integrated crash reporting tool for mobile apps \\
Instabug      & In-app bug and crash reporting tool \\
Bugsnag       & Error monitoring and crash reporting platform \\
\midrule
\multicolumn{2}{@{}l}{\textit{Advertising \& Monetization (9)}} \\
\midrule
Facebook      & Meta's ad targeting and conversion tracking SDK \\
AppLovin      & Mobile ad network and monetization platform \\
TikTok/Pangle & TikTok's ad network and audience network (Pangle) \\
Mintegral     & Programmatic in-app advertising and monetization SDK \\
AdMob         & Google's mobile in-app advertising network \\
Unity         & Unity's in-app advertising network (Unity Ads) \\
Fyber         & Mobile ad mediation and exchange platform \\
BidMachine    & Programmatic in-app advertising exchange \\
InMobi        & Mobile advertising network \\
\midrule
\multicolumn{2}{@{}l}{\textit{Push Notifications (1)}} \\
\midrule
OneSignal     & Push notification and in-app messaging service \\
\bottomrule
\end{tabularx}
\caption{Brief description of the trackers found by our analysis.}
\label{tab:tracker_description}
\end{table}

\descr{Third-Party Domains (mitmproxy/PCAPdroid).}
\label{sec:dyn-observed}
Finally, we capture traffic via mitmproxy and PCAPdroid for 20/25 apps.
Every app contacts at least one tracker-associated domain during app usage.
Table~\ref{tab:wireshark} lists the domains contacted by each app.
The most frequent is Firebase, in 16 (80\%) apps, followed by Crashlytics (12), AppsFlyer (9), and Facebook (9).
These services differ in their purpose.
Table~\ref{tab:wireshark} lists 77 app-service pairs.
Using the categories presented in Table~\ref{tab:tracker_description}, 32 (42\%) are advertising or attribution services, and 14/20 apps contact at least one.
The rest are analytics, crash reporting, push notifications, and similar operational services.
Talkie contacts the highest number of distinct advertising-related domains and, as discussed later, none of this is mentioned in its privacy policy.
Table~\ref{tab:tracker_description} also briefly describes each identified tracker.

\descr{AI and Cloud Providers (mitmproxy/PCAPdroid).}
\label{sec:dyn-ai}
Beyond trackers, we determine which and how apps reach AI and cloud services.
Rocky winds up being the most exposed: it calls {\it Amazon Polly} voice generator service\footnote{\url{https://polly.eu-central-1.amazonaws.com/v1/speech}} directly from the user device to synthesize each coaching message.
It actually sends each message text in cleartext and stores the resulting audio in an Amazon S3 bucket, with object keys embedding the spoken sentence.
None of this appears in Rocky's privacy policy: it mentions neither Amazon nor Polly, nor any speech synthesis, and refers to the app only in generic terms such as ``AI coaching.''
Polly requests are signed on the device with a permanent AWS access key rather than a temporary one, meaning the app contains valid AWS credentials.
Notably, MobSF's composite App Security Score does not capture this exposure: Rocky scores $56$, the joint-highest in our corpus, while shipping a permanent AWS access key in the APK.
This is why we do not report these scores as a security metric.

In contrast, Rosebud's journaling and reflection calls are handled on Rosebud's own servers rather than on the user device.
Therefore, we cannot observe model requests directly.
However, we do observe a provider manifest that the app downloads at runtime\footnote{\url{https://my.rosebud.app/static/providers.json}}, which lists production models it uses, including a Google model (Gemma).
Rosebud's policy names OpenAI, Anthropic, and Groq, though not Google.

\subsection{Policy Claims} 
\label{sec:policy-claims}
Next, we present results from analyzing apps' privacy policies with Gemini 3.1 Pro, following the pipeline presented in Section~\ref{sec:policy-method}.
We first validate extracted labels against manual annotation, including wherever the second model (GPT-5.4) disagrees, and quantify how residual errors affect our counts.
Next, we report what policies disclose about third-party AI providers.
Section~\ref{sec:disclosure-gap}, compares policy disclosures with apps' behaviors (from Section~\ref{sec:behavior-evidence}).

\descr{Extraction Accuracy.}
\label{sec:rq3}
We validate LLM-extracted labels against manual annotations of  24 privacy policies (Headspace and Headspace Care share one).
For each policy, the LLM produces five binary disclosure labels: camera/photo collection, microphone/audio collection, named AI/LLM providers, named third-party trackers, and advertising.
The extraction prompt (see Appendix~\ref{appendix:prompt}) is written with particular attention to camera and microphone categories, since these are most directly tied to sensitive data capture in the therapy context (Section~\ref{subsec:static}).
We extract the other three categories with the same prompt structure.
We then compare these labels to the corresponding manual annotations (performed by two authors) of the policy text.
We do this independently for each LLM.
Table~\ref{tab:llm-validation-combined} reports the alignment of Gemini 3.1 Pro and GPT-5.4 with manual annotation.

\begin{table}[t]
\centering
\small
\setlength{\tabcolsep}{4pt}
\begin{threeparttable}
\begin{tabular}{lcccccc}
\toprule
& \multicolumn{3}{c}{\textbf{Gemini 3.1 Pro}} & \multicolumn{3}{c}{\textbf{GPT-5.4}} \\
\cmidrule(lr){2-4} \cmidrule(lr){5-7}
\textbf{Category} & \textbf{Prec} & \textbf{Rec} & \textbf{F1} & \textbf{Prec} & \textbf{Rec} & \textbf{F1} \\
\midrule
Camera/Photo       & 1.00 & 1.00 & 1.00 & 1.00 & 1.00 & 1.00 \\
Microphone/Audio   & 0.92 & 1.00 & 0.96 & 0.92 & 1.00 & 0.96 \\
Named AI Providers & 1.00 & 1.00 & 1.00 & 1.00 & 1.00 & 1.00 \\
Named Trackers & 1.00 & 0.93 & 0.97 & 1.00 & 0.93 & 0.97 \\
Advertising        & 1.00 & 0.91 & 0.95 & 0.96 & 1.00 & 0.98 \\
\midrule
\textbf{Macro Average} & \textbf{0.98} & \textbf{0.97} & \textbf{0.97} & \textbf{0.97} & \textbf{0.99} & \textbf{0.98} \\
\bottomrule
\end{tabular}
  \begin{tablenotes}[flushleft]
\footnotesize
    \item {\em Abbreviations:} Prec = Precision, Rec = Recall, F1 = F1 score.
    \item[]\vspace{-5pt}\hspace*{-\labelsep}\rule{\linewidth}{0.7pt}
  \end{tablenotes}
  \end{threeparttable}
\caption{LLM extraction accuracy against reconciled manual annotation for Gemini 3.1 Pro and GPT-5.4 (24 policies).}
\label{tab:llm-validation-combined}
\end{table}

Gemini 3.1 Pro achieves a macro-average F1 of $0.97$ (precision $0.98$, recall $0.97$), and GPT-5.4 achieves $0.98$ (precision $0.97$, recall $0.99$).
Both models agree with the manual annotation on every camera/photo and named-AI-provider label, and they disagree with it on only 7 of 240 labels in total.
Both models flag the same five policies as naming AI providers.
This bolsters our confidence that the extraction approach is not specific to one model.

Both models mark NeuroFit as disclosing audio collection, quoting text that appears in the Terms of Use, which NeuroFit publishes on the same page as its privacy policy, rather than in the policy itself.
Both models also miss the one tracker that Stoic names: Amplitude appears only in the cookie-preference table served with its policy, not in the policy text.
The remaining errors concern advertising.
GPT-5.4 marks Rocky as disclosing advertising based on marketing e-mails and boilerplate about device advertising settings, but neither states that data is shared for advertising.
Gemini 3.1 Pro misses advertising for Wysa and Stoic: both policies deny selling or sharing data with advertisers, but Wysa also states that it uses Meta Ads Manager to create and measure ads, and Stoic's cookie table lists cookies ``used for targeted advertising,'' contradicting itself. %

\descr{Impact on Disclosure Counts.}
Across the 240 label decisions (24 policies $\times$ 5 categories $\times$ 2 models), the LLM classifiers disagree with our reconciled manual labels in 7 cases: 3 false positives and 4 false negatives, in the audio, named-tracker, and advertising categories.
Hence, they could change at most one camera/microphone disclosure decision (NeuroFit's microphone) and one named-tracker decision (Stoic's Amplitude), and no AI-provider label.
The gap we report later between tracker disclosure ($15$\% of detected items disclosed) and permission disclosure ($61$\%) in Section~\ref{sec:disclosure-gap} is far too big to be an extraction error.
This settles our third axis: extraction errors are small and do not materially shift the disclosure counts reported in Section~\ref{sec:disclosure-gap}.

\descr{AI Provider Disclosures.} 
We find that 14/25 apps ($56$\%) reference sending user data to third-party AI providers in their privacy policies -- see Table~\ref{tab:ai_processing}.
More precisely, $5$ apps name specific providers, while $9$ use generic language, such as ``AI services'' or ``large language models,'' without naming the entity that receives user data.
Notably, 3 of the 11 apps whose policies do not reference third-party AI processing (Talkie, Ash, and Neurofit) offer an AI chatbot as their primary interface.
Naturally, the wording and level of detail vary significantly from app to app.
For instance, Rosebud's policy mentions three AI providers:
{\it ``OpenAI, Anthropic, and Groq: We use these services for advanced computational tasks, including but not limited to, artificial intelligence operations.''}
Worryingly, users who record voice journals about depression or trauma might not realize that their entries are processed by three separate external entities.
Across the corpus, named providers span OpenAI, Anthropic, Groq, and Google Gemini, e.g., Stoic names OpenAI, Anthropic, and Google Gemini, while Aura names Claude.
Rosebud's policy names OpenAI, Anthropic, and Groq, but not Google, even though mitmproxy recorded the Rosebud app fetching a provider manifest that lists a Google model (Gemma) among its production models.

Policies also vary on whether user content trains AI models.
Noah, Pi, 7~Cups and Youper explicitly mention doing it. %
Stoic and Wysa explicitly rule out training use; BetterUp rules it out for general-purpose models, and Replika -- for third-party models, while training its own safety algorithms on anonymized portions of messages.
The remaining apps that mention AI (Rocky, Aura, Yana, Headspace, Rosebud, Clarity, Hapday) say nothing about training.

\begin{table}[t]
\centering
\small
\setlength{\tabcolsep}{3.5pt}
\begin{threeparttable}
\begin{tabular}{@{}llc@{\hskip 0.6em}llc@{}}
\toprule
\multicolumn{3}{c}{\textbf{Named Provider}} & \multicolumn{3}{c}{\textbf{Generic Language}} \\
\cmidrule(lr){1-3} \cmidrule(lr){4-6}
\textbf{App} & \textbf{Provider} & \textbf{Data} & \textbf{App} & \textbf{Term} & \textbf{Data} \\
\midrule
Rosebud & Open, Antr, Groq   & Journal   & BetterUp  & AI  & User \\
Aura    & Antr              & Msgs      & Clarity & AI  & User \\
Stoic   & Open, Antr, Gem    & User      & Hapday & AI  & User \\
Yana    & Open              & User      & Headspace & AI & User \\
Youper  & Open              & User      & Noah      & AI  & Emot. \\
        &                  &           & Pi        & LLM & Msgs \\
        &                  &           & Replika   & AI  & Msgs \\
        &                  &           & Rocky     & AI  & Msgs \\
        &                  &           & Wysa      & LLM & Msgs \\
\bottomrule
\end{tabular}
  \begin{tablenotes}[flushleft]
\footnotesize
    \item %
    \hspace{-0.1875cm} {\em Abbreviations:} Open = OpenAI, Antr = Anthropic, Gem = Gemini; Journal = journal entries, Msgs = messages, User/Emot. = user/emotional data.
    \item[]\vspace{-5pt}\hspace*{-\labelsep}\rule{\linewidth}{0.7pt}
  \end{tablenotes}
  \end{threeparttable}
\caption{AI provider references in privacy policies. }
\label{tab:ai_processing}
\end{table}

Five apps naming specific providers make voluntary admissions in potentially legally binding documents, although we cannot verify that they actually share user data.
Nine apps use only generic language, thus leaving users unable to determine exactly who receives their data, a disclosure gap with direct implications under CCPA \S1798.110(c)(4)~\cite{ccpa}, which requires businesses to disclose ``the categories of third parties to whom the business discloses personal information.''
Regulatory implications are considered in Section~\ref{sec:ccpa}.

\subsection{Disclosure Gap}
\label{sec:disclosure-gap}
We now compare what each app does (Section~\ref{sec:behavior-evidence}) with what its privacy policy says (Section~\ref{sec:policy-claims}).
This allows us to determine how often each app excludes an observed practice from its policy across the app corpus and three analysis categories: trackers, camera/microphone permissions, and AI providers.
Table~\ref{tab:master-disclosure} reports the counts of detected/disclosed items for every app and the gap between them.

\begin{table}[t]
\centering
\small
\setlength{\tabcolsep}{5pt}
\begin{tabular}{lrr@{\hspace{8pt}}rr}
\toprule
& \multicolumn{2}{c}{\textbf{Trackers}} & \multicolumn{2}{c}{\textbf{Permissions}} \\
\cmidrule(lr){2-3} \cmidrule(lr){4-5}
\textbf{App} & Detected & Disclosed & Detected & Disclosed \\
\midrule
7~Cups          &   3 &   0 &   2 &   2 \\
Ash             &   4 &   0 &   2 &   1 \\
Aura            &  11 &   1 &   2 &   0 \\
BetterUp        &   4 &   0 &   2 &   2 \\
CBT Guide &   3 &   1 & N/A & N/A \\
Clarity         &   5 &   1 &   1 &   0 \\
Hapday          &   7 &   0 &   2 &   2 \\
Happify         &  11 &   2 &   1 &   1 \\
Headsp. Care    &   5 &   0 &   2 &   2 \\
Headspace       &   4 &   0 &   2 &   2 \\
Limbic          &   2 &   2 &   1 &   0 \\
MindDoc         &   5 &   4 & N/A & N/A \\
Neurofit        &   2 &   0 &   2 &   1 \\
Noah            &   2 &   0 &   2 &   1 \\
Pi              &   5 &   0 &   1 &   1 \\
Replika         &   6 &   0 &   2 &   2 \\
Rocky           &   4 &   1 &   1 &   0 \\
Rosebud         &   3 &   2 &   2 &   0 \\
Sensa           &   6 &   2 &   1 &   0 \\
Stoic           &   2 &   1 & N/A & N/A \\
Talkie          &  21 &   0 &   2 &   2 \\
Woebot          &   1 &   0 & N/A & N/A \\
Wysa            &   3 &   1 &   1 &   1 \\
Yana            &   5 &   0 &   1 &   0 \\
Youper          &   3 &   1 &   1 &   0 \\
\midrule
\textbf{Total} & \textbf{127} & \textbf{19} & \textbf{33} & \textbf{20} \\
\bottomrule
\end{tabular}
\caption{Gaps in tracker and permission disclosures.
24/25 apps have at least one tracker gap; 11/21 apps (52\%) declaring camera or microphone have a permission gap.}
\label{tab:master-disclosure}
\end{table}

\begin{figure}[t]
    \centering
    \includegraphics[width=0.9999\linewidth]{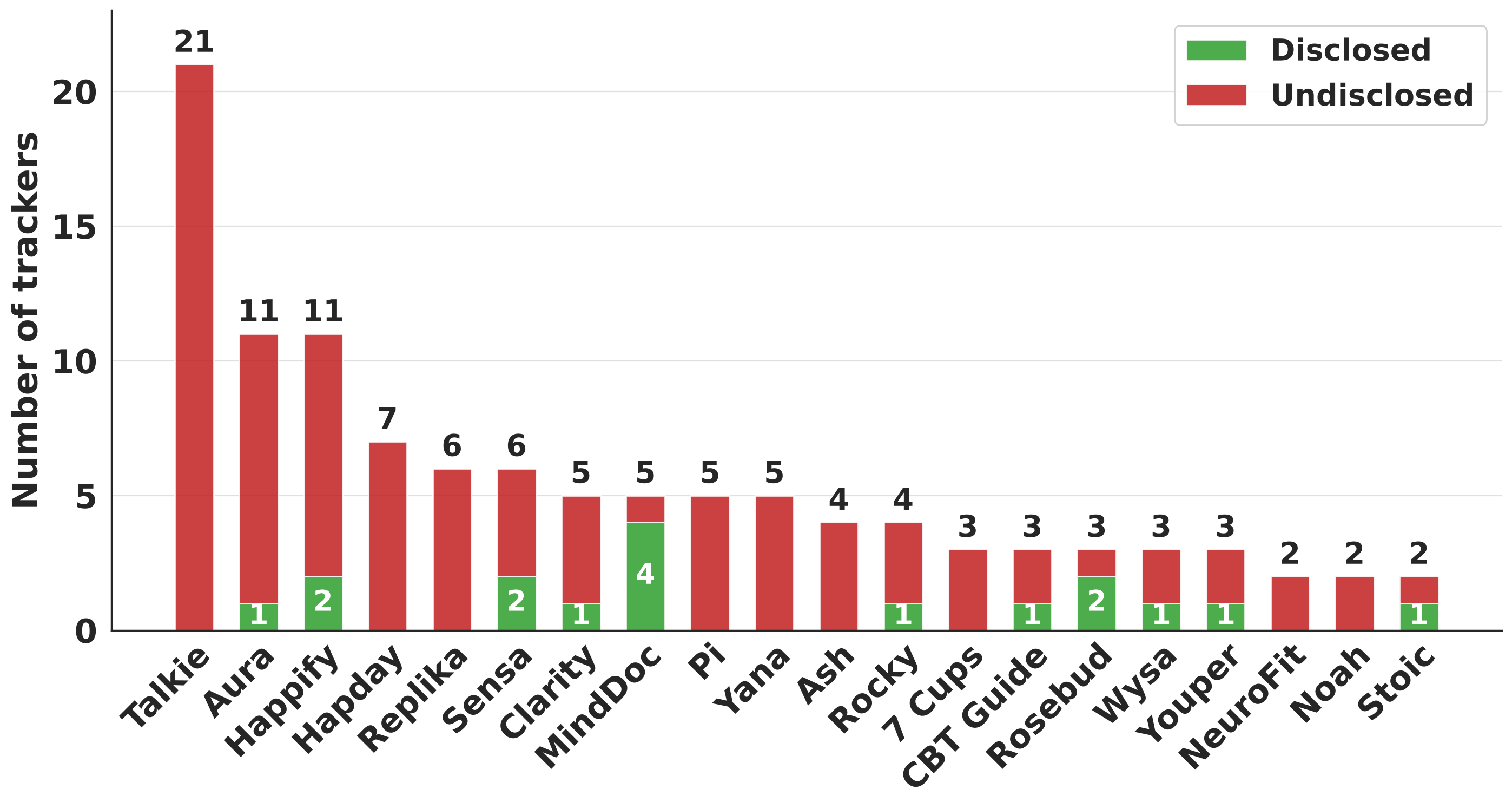}
    \caption{Tracker disclosure per app, for the 20 apps we instrument at runtime.
    Each tracker SDK detected by static analysis (MobSF) is classified by whether the privacy policy names it.}
    \label{fig:disclosure-gap}
\end{figure}

\descr{Coverage.}
Tracker counts in Table~\ref{tab:master-disclosure} come from static analysis and cover all 25 apps, and Figure~\ref{fig:disclosure-gap} plots the same counts for the 20 apps we could exercise (see Table~\ref{tab:coverage}).

\descr{Undisclosed Practices.}
\label{sec:rq1}
No app achieves full disclosure: all but Limbic embed at least one tracker SDK their privacy policy does not name (Limbic's policy names both of its trackers but not the \perm{record\_audio} permission it declares), and $13$ apps ($52$\%) name no embedded trackers at all.
For camera and microphone, of $21$ apps that request either permission, $11$ ($52$\%) fail to disclose at least one, while $10$ ($48$\%) disclose every one they declare.
For instance, although $4$ apps (Wysa, Youper, Happify, and Aura) include statements on their initial sign-in screens assuring users that their data is ``private'' or ``confidential,'' they embed multiple undisclosed third-party trackers.

\descr{Tracker Disclosure Gap.}
For each app, we compare trackers detected using both static and dynamic analyses with those explicitly named in its privacy policy.
Across 20 apps with both static and dynamic data, $18$ ($90$\%) fail to disclose at least half of detected trackers.
The most extreme case is Talkie: its APK embeds 21 distinct trackers (including Adjust, AppLovin, AppsFlyer, Facebook, Google AdMob, Mintegral, TikTok/Pangle, and ironSource) while its privacy policy names none.
Under GDPR Article~13(1)(e)~\cite{gdpr}, controllers must disclose ``the recipients or categories of recipients of the personal data, if any.''
Generic statements such as ``we use analytics services'' are grossly inadequate if an app embeds 21 specific data recipients.
We do not claim that every third-party service is equally invasive: a crash reporter differs from an advertising network.
The disclosure duty, however, applies to all recipients of personal data regardless of purpose.

\begin{table}[t]
\centering
\small
\setlength{\tabcolsep}{6pt}
\begin{tabular}{@{}ll@{}}
\toprule
\textbf{App} & \textbf{Undisclosed Permission} \\
\midrule
\multicolumn{2}{@{}l}{\textit{Camera / Microphone (13)}} \\
\midrule
Ash      & \perm{camera} \\
Aura     & \perm{camera}, \perm{record\_audio} \\
Clarity  & \perm{record\_audio} \\
Limbic   & \perm{record\_audio} \\
Neurofit & \perm{record\_audio} \\
Noah     & \perm{camera} \\
Rocky    & \perm{record\_audio} \\
Rosebud  & \perm{camera}, \perm{record\_audio} \\
Sensa    & \perm{record\_audio} \\
Yana     & \perm{record\_audio} \\
Youper   & \perm{camera} \\
\midrule
\multicolumn{2}{@{}l}{\textit{Other Dangerous Permissions (10)}} \\
\midrule
Aura            & \perm{read/\allowbreak write\_calendar} \\
Headspace       & \perm{read/\allowbreak write\_calendar} \\
Headspace Care  & \perm{read/\allowbreak write\_calendar} \\
Talkie          & \perm{read/\allowbreak write\_calendar} \\
Rocky           & \perm{get\_accounts} \\
CBT Guide & \perm{write\_external\_storage} \\
Talkie          & \perm{read/\allowbreak write\_external\_storage} \\
Stoic           & \perm{read\_phone\_state} \\
Happify         & \perm{body\_sensors}, \perm{activity\_recognition} \\
Noah            & \perm{activity\_recognition} \\
\bottomrule
\end{tabular}

\caption{All 23 manually-verified permission-policy contradictions, grouped by data category.
\perm{read\_phone\_state} corresponds to device identifiers and \perm{body\_sensors}/\perm{activity\_recognition} to physical-activity data.}
\label{tab:contradictions}
\end{table}

\descr{Permission-Policy Contradictions.}
We identify $23$ contradictions between permissions and policies across $17$ apps, i.e., where Androguard confirms that a dangerous permission is declared in the manifest, but the policy-analysis pipeline (validated against manual annotation) finds no corresponding mention in the app's privacy policy.
We list camera/microphone contradictions first, because they are the most worrisome subset: these permissions enable direct sensing of the user's voice and likeness in the context of therapy interactions, a substantively more sensitive disclosure failure than, e.g., a missing mention of \perm{write\_external\_storage}.
For every dangerous permission an app declares, we check whether its policy discloses the corresponding data: camera/microphone access, device identifiers (\perm{read\_phone\_state}), calendar entries (\perm{read\_calendar}/\allowbreak\perm{write\_calendar}), on-device accounts (\perm{get\_accounts}), physical activity and sensor data (\perm{activity\_recognition}, \perm{body\_sensors}), and file or media access (\perm{read\_external\_storage}/\allowbreak\perm{write\_external\_storage}).
Table~\ref{tab:contradictions} groups the 23 contradictions by data category.

Our pipeline initially flags seven additional declarations, which we exclude after manual review: in each case, the policy does disclose the data, but in indirect language that the LLM extractor missed.
More precisely, Headspace and Headspace Care list ``audio recordings, voicemails, photographs,'' while Aura, Headspace, Headspace Care, and Replika disclose device-identifier collection in different phrasing, and Hapday discloses step and sleep tracking.

Among 21 apps that request camera or microphone permissions, 12 have a functional reason for the permission: video therapy (Headspace, Headspace Care, 7~Cups, BetterUp), voice journaling (Rosebud, Wysa), AI voice or chat (Replika, Talkie, Pi, Noah, Ash), and profile-photo upload (Youper).
8 of 12 disclose every camera or microphone permission they declare in their privacy policy.
Rosebud, Youper, Ash, and Noah are the exceptions: Rosebud uses \perm{record\_audio} for voice journaling and Youper uses \perm{camera} for profile photos, yet neither policy discloses it, while Ash and Noah disclose audio collection but not the \perm{camera} permission they also declare; all four appear among the camera/microphone contradictions in Table~\ref{tab:contradictions}.
The other 7 apps with contradictions (Aura, Rocky, Sensa, Yana, Clarity, Limbic, and NeuroFit) do not have a clear functional use for the permission they declare, and also do not disclose it.

\section{Related Work \label{sec:rel}}
We now review prior work on security/privacy analysis of Android apps, transparency and compliance with privacy policies, and privacy risks in health and AI companion apps.

\descr{Static Analysis.} 
Kollnig et al.~\cite{kollnig2021iphones} use static (and traffic) analysis to study third-party tracking across 24k iOS and Android apps, finding widespread tracking and unique user identifier sharing, including in apps aimed at children.
Wessels et al.~\cite{wessels2025hytrack} focus on persistent cross-context tracking that abuses the browser state shared with Custom Tabs and Trusted Web Activities, while Specter et al.~\cite{specter2025fingerprinting} map the market of SDKs that enable device fingerprinting.
Girish et al.~\cite{girish2025your} analyze 52 wireless-scanning SDKs, identifying cross-library sharing of location data and device identifiers.
We use static analysis %
not to measure tracking across the ecosystem, but to analyze the gap between the SDKs apps embed and what they account for in their privacy policies.

\descr{Dynamic Analysis.} 
Along with static analysis, Reardon et al.~\cite{reardon2019fifty} monitor runtime behavior and traffic across 250k versions of 88k Android apps, uncovering side and covert channels used by apps and third-party SDKs to access sensitive data without permission.
Razaghpanah et al.~\cite{razaghpanah2018apps} analyze real-world traffic from almost 15k apps collected via the Lumen Privacy Monitor, identifying 2k+ advertising and tracking services at the traffic level.
Dong et al.~\cite{dong2024exploring} discover covert third-party identifiers stored in external storage to bypass Android's identifier restrictions.
Weerasekara et al.~\cite{weerasekara2025tracking} combine static and dynamic analysis to identify cross-context tracking through Android WebViews, including cookie synchronization and fingerprinting.
Unlike these large-scale, mostly automated measurements, we exercise each app manually, on a physical device, bypass certificate pinning where feasible, and use the observed endpoints to corroborate statically detected SDKs and identify AI and cloud providers contacted at runtime.

\descr{Sensitive Apps.} 
Dong et al.~\cite{dong2022privacy} analyze privacy practices of period-tracking apps following Dobbs v.~Jackson~\cite{dobbs2022}, demonstrating the importance of domain-specific privacy analysis for apps that handle sensitive health data.
Huckvale et al.~\cite{huckvale2019assessment} intercept traffic from 36 depression-focused and smoking cessation apps, finding that 29 transmitted data to Google or Facebook services. %
Similar to our work, they study observed behavior%
, although only from intercepted traffic, while we add static analysis of embedded SDKs and declared permissions, as well as third-party AI providers.

In an upcoming study, Girish et al.~\cite{girish2027sweat} analyze 1,548 Android health and fitness apps and 11 integration platforms (e.g., Google Fit, Fitbit, Strava).
They find that OAuth-authorized integrations move health records between apps outside Android's permission model and that 57\% of apps send health data with persistent identifiers to third parties.
While they measure data flows, we examine whether privacy policies disclose it, focusing on mental health and coaching apps (whose users may share even more intimate content).

More closely related to our work, Iwaya et al.~\cite{iwaya2022privacy} analyze 27 Android mental health apps using static analysis (MobSF), dynamic traffic capture, and privacy-policy evaluation mapped %
to the LINDDUN threat taxonomy.
They %
document insecure cryptographic implementations, hard-coded secrets, plaintext storage, and leaked credentials.
Besides focusing on an older ecosystem (their study is from 2022), our work differs in several ways.
First,~\cite{iwaya2022privacy} does not compare, per app and per data category, observed behavior against policy disclosure, relying on off-the-shelf policy tools (CLAUDETTE at 78\% accuracy, PrivacyCheck at 60\%), whereas we validate LLM extraction (macro-average F1 of $0.97$--$0.98$).
Second, their study predates the generative-AI shift: we analyze third-party AI-provider disclosure, %
and training-use clauses, %
and find AI providers layered on a similarly invasive infrastructure (e.g., Firebase, Facebook) documented in prior work~\cite{iwaya2022privacy,huckvale2019assessment}.
Third, their corpus is de-identified, precluding accountability and independent reproduction, while our public artifact release can enable the enforcement analysis of Section~\ref{sec:ccpa}. %
Finally, their dynamic analysis runs on an Android emulator that could not exercise several apps, while %
we capture on a physical device and circumvent certificate pinning whenever feasible.

\descr{Privacy Policy Compliance.} 
Prior work has also examined whether app privacy policies accurately reflect actual data practices.
Okoyomon et al.~\cite{okoyomon2019ridiculousness} analyze 68,051 Android apps, identifying discrepancies between observed third-party transmissions and privacy-policy disclosures.
Andow et al.~\cite{andow2019policylint} detect internal contradictions within policies on Google Play and inconsistencies between policy claims and actual data flows~\cite{andow2020actions}.
Zimmeck et al.~\cite{zimmeck2017automated} combine machine-learning-based privacy policy analysis with static code analysis of 17,991 Android apps, demonstrating the viability of automated compliance detection.
Follow-up work~\cite{zimmeck2019maps} also finds that many app privacy policies are either absent or silent about the apps' actual practices.
Samarin et al.~\cite{samarin2023lessons} use an instrumented Android build to log traffic from 109 apps and compare it against both privacy-policy claims and developers' responses to CCPA ``right to know'' requests, finding that only 36\% of responding developers disclosed the CCPA-required categories of third-party recipients.
We adopt the same detected-vs-disclosed design, on a corpus small enough to validate the policy analysis manually for every app, and extend it to third-party AI providers and training-use clauses.

\descr{Platform Transparency.} 
Kollnig et al.~\cite{kollnig2021consent} find that 71.3\% of 1,297 Google Play apps contact tracker hosts at launch, yet only 9.9\% ask for consent, which they argue points to widespread violations of EU and UK privacy law~\cite{gdpr}. %
Mohamed et al.~\cite{mohamed2024attention} investigate Apple's App Tracking Transparency framework, finding that 59\% of ATT alerts across 4,000 iOS apps use dark patterns that mislead users into granting tracking permission, while Reyes et al.~\cite{reyes2018coppa} show that a majority of the $5{,}855$ children's apps they analyze collect and share personal information in ways that are potentially in violation of federal regulations protecting minors, such as COPPA.
More recently, Alomar et al.~\cite{alomar2025effect} analyze 7,377 child-directed Android apps, finding persistent issues with third-party SDKs and inaccurate Data Safety disclosures despite policy changes.
These studies assess consent flows, platform-level labels, and children's-privacy rules, whereas we assess each developer's own privacy policy against the app's observed behavior.

\descr{AI Companions.} 
Recent work examining risks in AI companion apps~\cite{brigham2026} provides a harm taxonomy for AI companion apps, e.g., sensitive data collection, anthropomorphism, engagement manipulation, and synthetic non-consensual imagery.
The authors also conduct a qualitative walk-through of 30 apps from a sample of 489, identifying risks to user privacy, though without inspecting the apps' APKs, manifest permissions, or runtime network traffic.
Kwesi et al.~\cite{kwesi2025exploring} interview 21 users of general-purpose LLM chatbots for mental health and find widespread misconceptions: participants conflated human-like empathy with human-like accountability, mistakenly believed their disclosures were covered by HIPAA-equivalent protections, and undervalued emotional disclosures relative to financial or location data.
Finally, Oliveira et al.~\cite{oliveira2027prompt} recently investigate privacy and tracking practices in web and mobile conversational AI services, whereas we focus on mental health and life-coaching Android apps and on the AI providers they integrate.

\section{Discussion}\label{sec:discussion}
\subsection{Regulatory Implications} \label{sec:ccpa}
\noindent{\bf United States.} The California Consumer Privacy Act (CCPA)~\cite{ccpa} requires businesses to disclose %
``the categories of third parties to whom the business discloses personal information'' (\S1798.110(c)(4), implemented by \S1798.130(a)(5)(B)(iv)).
CCPA stops short of requiring that recipients be named individually, a limitation that, according to some prior work~\cite{samarin2023lessons}, should be addressed. However, the categories it requires disclosure of must still be meaningful.
It is unclear whether the term ``service providers'' meaningfully describes entities, such as Facebook, AppsFlyer, Google AdMob, Amplitude, Mixpanel, and 15 others, each with its own data practices.

The disclosure failures we document raise significant concerns under U.S.\ consumer-protection law, particularly when apps make explicit promises of confidentiality that may not align with their observed data practices.
Section~5 of the Federal Trade Commission (FTC) Act (15 U.S.C.\ \S 45) prohibits ``unfair or deceptive acts or practices in or affecting commerce''~\cite{ftcact}, and the FTC has used this authority to act against mental-health services that share user data with advertisers while assuring users of confidentiality.

In 2023, the FTC took action against BetterHelp (an online therapy provider matching users with licensed therapists; not in our corpus), which agreed to pay \$7.8M for sharing sensitive mental-health data with Facebook, Snapchat, Criteo, and Pinterest after assuring users their data would remain confidential~\cite{ftc2023betterhelp}.
Several apps in our dataset make similar ``private'' or ``confidential'' assurances at the start of use (see Section~\ref{sec:initial-collection-results}) while embedding tracker SDKs which are not cited in their privacy policies.

\descr{GDPR.}
Under GDPR Article~13(1)(e)~\cite{gdpr}, data controllers must provide data subjects with ` `the recipients or categories of recipients of the personal data, if any,'' and must do so `` at the time when personal data are obtained.''
The Article~29 Working Party's guidelines~\cite{wp29} state that, in practice, this ``will generally be the named recipients, so that data subjects know exactly who has their personal data,'' and that where a controller opts to disclose categories, ``the information should be as specific as possible by indicating the type of recipient.''
Again, generic terms like third parties or service providers do not satisfy this criterion.
Alas, nearly all the apps we instrument (18/20) fail to disclose at least half of their embedded trackers in their privacy policies, and about one-third (7/20) use exclusively generic language while embedding 3 or more specific trackers.
In particular, the popular chatbot app Talkie embeds 21 distinct trackers while naming zero in its privacy policy.

\descr{The Replika Precedent.}
In April 2025, the Italian data protection authority issued a \texteuro5M fine to Replika's developer for GDPR violations, finding privacy notices and information practices inadequate and age-verification mechanisms deficient~\cite{replika-fine}.
Our analysis identifies similar issues across the therapy app ecosystem.
About half of the apps (13/25) name none of the trackers we detect, and even those that name some disclose only a fraction, such as Aura (1/11) and Happify (2/11).
Replika itself names none of the 4 trackers our dynamic analysis uncovers. 
While the Replika case suggests that regulators are beginning to scrutinize this domain, our analysis indicates widespread exposure to similar enforcement risk.

\descr{SDK Vendors.} One plausible explanation is that developers do not know what an embedded SDK collects.
However, vendors of widely embedded SDKs document this information.
By default, the Google Analytics for Firebase SDK collects the Android advertising ID along with screen views and sessions~\cite{ga-firebase-data}, the Facebook SDK ``automatically logs common mobile events from your app like app installs and app launches''~\cite{meta-sdk-gdpr}, and AppsFlyer's terms name advertising IDs, IP addresses, and device IDs as end-user data~\cite{appsflyer-terms}.
The terms also require developers to tell their users.
Google requires developers to post a privacy policy and to disclose ``the use of Google Analytics, and how it collects and processes data''~\cite{ga-terms} and, for Crashlytics, a privacy policy stating that user information is ``disclosed to and processed by third party providers like Google''~\cite{crashlytics-terms}.
Meta and AppsFlyer have comparable notice duties~\cite{meta-business-tools,appsflyer-terms}.
The omissions we report could be interpreted as a breach of terms they accepted, beyond any GDPR/CCPA issues.
Moreover, Meta, AppsFlyer, Amplitude, and Sentry prohibit or restrict sending health information through their services~\cite{meta-business-tools,appsflyer-terms,amplitude-terms,sentry-terms}.
Even though we cannot determine whether usage events from a therapy app fall under these restrictions, this issue deserves scrutiny.

\subsection{Data Deletion}
GDPR Article~17~\cite{gdpr} and CCPA \S1798.105~\cite{ccpa} grant users the right to request deletion of personal data.
This motivates examining each app's stated deletion mechanisms (if any) and the means of exercising them.
We also note the email addresses that app users can contact; see Table~\ref{tab:deletion-contacts} in Appendix~\ref{app:contact}.

Overall, 8/25 (32\%) of apps offer in-app deletion, 22/25 (88\%) accept email deletion requests, and 2/25 (8\%) direct users to a web portal.
Note that counts overlap, since several apps offer multiple ways to delete data.
For 15/25 (60\%), email is the only option, which forces deletion-minded 
users to leave an external trail that they used a therapy app and want their data removed.
One app (CBT Guide) does not mention any deletion mechanism, which may violate both GDPR and CCPA.

Also, even if data deletion is possible, it might not be complete since data previously sent to third-party AI providers may be embedded in their model weights, which cannot be selectively erased~\cite{cooper2024unlearning}.

\subsection{AI Training}
Sending data to third-party AI providers also raises serious concerns.
First, every additional provider expands the attack surface for data breaches.
For instance, Rosebud's use of three providers means a user's journal entries may potentially exist on four separate systems, each with its own security practices.
Second, data used to train AI models may never be fully removed.
While sending data to a third-party AI provider does not imply training use, several apps in our dataset state in their policies that they indeed use user content for training.\footnote{For example, Noah mentions user content to ``train, refine, and improve'' models, Pi to ``develop and train our AI models/large language models,'' and Youper may ``analyze the content of conversations to understand opportunities for new features, model training, and fine-tuning.''}
As discussed, e.g., in~\cite{cooper2024unlearning}, specific information cannot be reliably removed from a trained model short of costly retraining.
Thus, even if a user deletes their account, their mental health disclosures may already be embedded in model weights or training datasets that cannot be selectively erased.
Moreover, 9 apps in our corpus reference AI providers without naming them, making it impossible for users to know where their data is sent and how to request its deletion.

\subsection{Why Undisclosed Trackers Cause Harm}
\noindent{Mental health status inference through cross-app tracking.} Advertising SDKs (e.g., Facebook) transmit a user's Google Advertising ID (AAID) along with app-usage events to their servers.
Because the same AAID is shared across every app on a device, a tracker embedded in both a therapy app and a shopping app can link activities between them.
Kollnig et al.~\cite{kollnig2021consent} show that, although most Android apps engage in third-party tracking, fewer than $3.5$\% obtain valid consent (i.e., offer a genuine option to refuse) under the ePrivacy Directive~\cite{edps2026eprivacy}.
This means this cross-app linking happens silently and at a massive scale.
In practice, an ad network that receives SDK calls from Replika, Wysa, and BetterUp can determine that a user is seeking mental health support without seeing a single conversation.
In fact, Gonz\'alez Caba\~nas et al.~\cite{cabanas2021facebook} show that Facebook assigns health-related ad-interest labels, such as ``suicide prevention'' and ``oncology,'' to users, based on inferred behavior, in the vast majority of cases without explicit consent.
In our dataset, Facebook's SDK appears in 9 of 25 apps, and 6 of these (Clarity, Hapday, Replika, Rocky, Talkie, and Yana) do not name Facebook in their privacy policy.

\descr{Implications for the data broker market.}
Data brokers are already advertising the ability to sell sensitive mental health information.
In 2023, Kim~\cite{kim2023databrokers} found 11 brokers willing to sell data on Americans' mental health conditions (including depression, anxiety, bipolar disorder, and PTSD) for as little as \$0.06 per record at volume. %
Some brokers also offered names, home addresses, and income.
Most mental health apps fall outside HIPAA because they are not operated by covered entities~\cite{kim2023databrokers}, leaving no comprehensive federal privacy law governing how data collected through embedded trackers flows to data brokers -- only after-the-fact FTC enforcement against deceptive practices~\cite{ftcact}.
The trackers we detect (e.g., Facebook, AppsFlyer, and Amplitude) feed into the same advertising and data ecosystem that brokers draw from, raising serious concerns about a new stream of sensitive data entering the data broker ecosystem.

\subsection{Ethical Considerations}
This work analyzed publicly available Android apps and their public privacy policies.
It did not involve human participants, collection of user-generated content, or interactions with real users.
Therefore, it is considered exempt by our Institutional Review Board (IRB).
While our findings identify undisclosed trackers and permission-policy contradictions in named commercial apps, we do not exploit these gaps or intercept real user traffic.

\descr{App and Policy Data.}
We downloaded APKs directly from Google Play for 22 of the 25 apps using an authenticated apkeep client, and obtained the remaining three (Limbic, Youper, and Happify) from public mirrors (APKPure, APKMirror, APKCombo, and AndroZoo).
We retrieved each app's privacy policy from its publicly linked Google Play URL on 27 July 2026.
We retrieved one policy (NeuroFit) from an Internet Archive snapshot dated 1 May 2026 because its live page was unreachable in July; its effective date (September 2023) is unchanged from the live version.
Downloads were authenticated with a dedicated account created for this study, and no personal data was entered.

\descr{Responsible Disclosure.}
On November 15, 2026, we notified Rocky that its Android app embeds a long-term AWS access key (in v2.1.15 and v2.1.20), and recommended revoking it and issuing short-lived credentials from the developer's backend.
We did not use, retain, or share the key. %
On the same date, we also notified the developers of all 25 apps of the findings.

As of November 19, three developers replied.
(1) Noah stated that the app does not use the camera or activity-recognition permissions it declares and that its developers had been asked to update the policy and the declared permissions accordingly.
(2) Limbic confirmed that its privacy policy names both of its embedded tracker SDKs, and stated that the app neither uses the microphone nor asks users for access to it: \perm{record\_audio} is merged into the app's manifest at build time from the Intercom SDK it uses for in-app support, and that they will remove the permission in an upcoming release.
(3) Hapday acknowledged that its privacy policy should have named the third-party services we reported and stated that, on November 18, 2026, it updated the policy to name the analytics and attribution services it uses (including AppsFlyer and Google Tag Manager) and its AI providers (including Anthropic, Google, and OpenAI), and to explain what data each service may receive and why.

\descr{Apps Before Account Creation or First Use.}
To document data collected before account creation or first use, we manually installed and walked through each app using a dedicated research account using a pseudonymous identity.
Apps requiring payment information for free-trial access were enrolled using a dedicated institutional research payment method.
No real user data was entered beyond the minimum necessary to access app functionality.

\section{Conclusions \& Future Work}\label{sec:conclusion}
This paper presented a measurement study of 25 popular Android therapy and life-coaching apps, comparing privacy policy disclosures against static analysis and network traffic observations.
Our findings reveal consistent transparency failures across these apps: 90\% of the apps we instrument fail to disclose at least half their embedded trackers, and 56\% disclose sending user data to third-party AI providers.

While we rely on established tools, such as MobSF and Androguard, this study goes beyond prior work in two ways.
First, it extends per-app comparisons between observed behavior and policy disclosures to third-party AI providers and training-use clauses. In contrast, earlier studies of mental health apps~\cite{iwaya2022privacy} were conducted before the advent of generative AI and do not cover this topic. 
Second, it shows that capturing traffic on a physical device, with certificate pinning bypassed, exposes practices that neither static analysis nor privacy policies reveal, e.g., Rocky's calls to Amazon Polly signed with a permanent AWS key, and a Google model in Rosebud's provider list not in its policy.

The implications of our work are not merely technical. 
Vulnerable users seeking help disclose highly sensitive thoughts (such as suicidal ideation, various fantasies, as well as histories of abuse and trauma) to an app which they believe to be both private and secure. 
Such users might not make informed decisions if they are unaware of exactly what data is collected, where, how, and for how long it is stored, as well as with whom it is shared. 
We believe that therapy apps warrant the same standard of confidentiality obligations as the human therapists they increasingly substitute for.
Rather than deliberate wrongdoing (recall that we found no evidence of malicious intent), the same patterns across 25 widely installed apps suggest a systemic problem: the therapy-app industry lacks a standard for handling sensitive data.
This motivates a code of ethics for app-based therapy (e.g., auditing third-party libraries, disclosing trackers, and keeping privacy policies current), perhaps starting as voluntary guidance and, over time, morphing into laws and regulations.

\descr{Limitations \& Future Work.} %
Our study considers only Android-based therapy apps, and iOS versions may involve different trackers, permission models, and SDKs.
Also, our keyword search and popularity filtering %
might miss apps targeting specific disorders or practices such as Dialectical Behavior Therapy (DBT).
Our corpus covers widely installed apps, %
and a broader corpus is left to future work.

Five apps (Woebot, Headspace, Headspace Care, BetterUp, and Limbic) required organizational or invitation codes; thus, our analysis likely underestimated their tracker activity.
Also, static analysis may over-approximate runtime behavior, since an embedded tracker SDK might not be activated in every session, and it could miss obfuscated SDKs; we mitigate this by manually verifying tracker classifications and corroborating them with the domains contacted at runtime.
Our dynamic analysis did not isolate which specific actions (e.g., sending a chatbot query) trigger which trackers, which we plan to investigate in future work.

Furthermore, our AI-provider analysis would miss calls from apps' backend servers; %
if the downstream provider were visible, we could confirm it directly (e.g., Rocky calling Amazon Polly, Rosebud's manifest naming a Google model), but we cannot claim a complete account of every provider.
Finally, we measured declared rather than exercised permissions, focusing on the former as the disclosure target for the reasons discussed in Section~\ref{subsec:static}, while leaving the latter to future work.
\bibliographystyle{abbrv}
\bibliography{references}

\appendices

\section{App Corpus}
\label{appendix:app-details}
Table~\ref{tab:app-overview} describes the \dataset{} apps we study, grouped by primary function and ordered by popularity within each group.
For each app, we report how widely it has been installed, where and when it was released, what service it provides to the users, and how one can access and/or pay for it.
This information comes from each app's Google Play listing and the developer's marketing material, as observed during July 2026.
Although we assign each app a single primary function, most offer several, as shown in Figure~\ref{fig:app-categories}.

\begin{table*}[p]
\centering
\footnotesize
\hyphenpenalty=10000 \exhyphenpenalty=10000
\setlength{\tabcolsep}{3.5pt}
\begin{threeparttable}
\begin{tabularx}{\textwidth}{@{}p{1.27cm}rr>{\hsize=1.4\hsize\raggedright\arraybackslash}X>{\hsize=0.6\hsize\raggedright\arraybackslash}X@{}}
\toprule
\textbf{App} & \textbf{\#DLs} & \hspace{-0.15cm}\textbf{Country~Yr} & \textbf{Description} & \textbf{Access \& Approx.~ Pricing} \\
\midrule
\multicolumn{5}{@{}l}{\textbf{\em AI Chatbot (11)}} \\[1ex]
Replika$^*$ & 10M+ & US~2017 & Personalizable AI companion; user shapes avatar, personality, and relationship type over time & Freemium; \$19.99/mo--\$299.99 lifetime \\[12pt]
Talkie & 10M+ & SG~2023 & Browse and roleplay with AI character personas; community-published characters & Free with ads; \$9.99/mo ad-free \\[5pt]
Yana & 10M+ & MX~2020 & Spanish-first emotional-support chatbot with self-help tools; premium unlocks unlimited messages & Freemium; \$11.99/mo--\$149 lifetime \\[12pt]
Wysa & 1M+ & IN~2016 & CBT/DBT chatbot with self-help library (meditation, breathing); premium adds human coach messaging & Freemium; \$19.99/yr \\[12pt]
Youper$^\dagger$ & 1M+ & US~2015 & AI chatbot for CBT-based mood management; tracks moods, feelings, and thoughts & 7-day trial, then \$69.99/yr \\[5pt]
Pi$^*$ & 500K+ & US~2023 & Conversational ``therapyGPT''-style assistant from Inflection AI & Free \\[5pt]
Woebot$^\ddagger$ & 500K+ & US~2018 & CBT chatbot, originally direct-to-consumer; now provider-gated since June 2025 & Provider or university invite only \\[5pt]
Clarity & 500K+ & US~2014 & CBT courses, AI-powered chat, daily check-ins for emotional patterns & Subscription; 7-day trial, then \$59.99/yr \\[12pt]
Noah & 100K+ & SG~2018 & AI therapist chatbot with voice-call option & Free trial, then \$10/mo or \$40/yr \\[5pt]
Rocky & 100K+ & GB~2020 & AI life-coaching chatbot; modules on relationships, productivity, networking & Free; \$9.99/mo, teams \$19.90/seat \\[5pt]
Ash & 50K+ & US~2024 & AI mental-health chatbot with guided ``paths'' for anxiety and other topics & Free (18+ only) \\
\midrule
\multicolumn{5}{@{}l}{\textbf{\em CBT/Self-Help (3)}} \\[1ex]
Happify$^{\dagger*}$ & 500K+ & US~2015 & Self-guided learning ``tracks'' grounded in positive psychology; no chatbot, no therapists & Freemium; \$14.99/mo--\$139.99/yr \\[12pt]
Sensa & 100K+ & LT~2021 & Self-paced daily lessons and journaling for stress and mood; coach app, no AI chatbot & Subscription; \$29.99--\$59.99 \\[5pt]
CBT Guide & 100K+ & US~2012 & Depression screening, self-tests, mood logging & Free; \$8.99 once to remove ads \\
\midrule
\multicolumn{5}{@{}l}{\textbf{\em Meditation (4)}} \\[1ex]
Headspace$^*$ & 10M+ & US~2012 & Guided meditations, courses, ``Ebb'' AI companion, on-demand text/video chats with certified coaches, and licensed therapy/psychiatry & Freemium; \$12.99/mo--\$69.99/yr \\[12pt]
Aura & 1M+ & US~2017 & Personalized meditation, sleep stories, life coaching, hypnosis, ASMR & Freemium; \$41.99--\$129.99/yr \\[5pt]
Hapday & 10K+ & AE~2025 & Life-coach app with breathing exercises, mood tracker, and 1~hr/day free AI coaching & Freemium; \$7.99/mo--\$29.99/yr \\[5pt]
Neurofit & 10K+ & US~2022 & Nervous-system regulation exercises, daily check-ins, AI breakthrough coaching & 3-day trial, then \$19.99/mo--\$124.99/yr \\[5pt]
\midrule
\multicolumn{5}{@{}l}{\textbf{\em Mood/Journal (3)}} \\[1ex]
MindDoc & 1M+ & DE~2017 & Mood tracking, journaling, courses based on cognitive-behavioral and mindfulness approaches. No chatbot in free version; AI used for personalization & Freemium; \$4.49/mo--\$69.99/yr \\[12pt]
Stoic & 100K+ & US~2020 & Journaling, breathing exercises, guided meditations; premium adds AI-powered journaling insights & Freemium; \$39.99--\$99.99/yr \\[5pt]
Rosebud & 50K+ & US~2024 & AI journal with chatbot that learns from entries to give personalized prompts & Subscription; \$12.99/mo \\
%\addlinespace[10pt]
\midrule
\multicolumn{5}{@{}l}{\textbf{\em Teletherapy/Coaching (4)}} \\[1ex]
7~Cups$^*$ & 1M+ & US~2014 & 24/7 chat with trained volunteers, ``Noni'' chatbot, paid text/audio/video therapy with licensed therapists & Freemium; free peer chat to \$299/mo therapy \\[12pt]
Headspace Care~ & 100K+ & US~2012 & Employer- or insurer-provided mental health platform; ``Ebb'' AI companion plus video sessions with licensed therapists and psychiatrists & Employer, insurer, or school code only \\[12pt]
BetterUp$^*$ & 50K+ & US~2016 & Executive and life coaching with both AI Coach and human-coach sessions; mostly employer-distributed & Employer-sponsored or individual \\[5pt]
Limbic$^\ddagger$ & 10K+ & GB~2019 & AI chatbot used as a clinical intake/triage layer; produces a clinical report for an in-person clinician (NHS-facing) & Clinician invite only \\
\bottomrule
\end{tabularx}
  \begin{tablenotes}[flushleft]
\footnotesize
    \item \hspace{-0.12cm} {\em Abbreviations:} ASMR = Autonomous Sensory Meridian Response, CBT/DBT = Cognitive/Dialectical Behavior Therapy, NHS = UK's National Health Service.
    %\item 
    {\em Countries:} AE = United Arab Emirates, DE = Germany, GB = United Kingdom, IN = India, LT = Lithuania, MX = Mexico, SG = Singapore, US = United States.
    \item[]\vspace{-5pt}\hspace*{-\labelsep}\rule{\linewidth}{0.7pt}
  \end{tablenotes}
  \end{threeparttable}
\caption{The \dataset{} apps in the corpus, grouped by primary category and sorted by downloads within each group.
``\#DLs'' is the Google Play download band, ``Country'' the developer's country of registration, and ``Year'' the year the app first appeared on Google Play.
``Description'' summarizes the primary user-facing functionality, ``Access'' indicates how a new user obtains the app, and ``Pricing'' lists the published consumer plans at the time of data collection.\\[0.5ex]
Apps marked with $\dagger$ were de-listed from Google Play after our initial collection period, those with $\ddagger$ required invitation codes and were thus analyzed with limited functionality. Apps marked with $^*$ also offer a web interface alongside the Android app.\\[0.5ex]}
\label{tab:app-overview}\label{tab:app-details}
\end{table*}

\section{LLM Extraction Prompt}
\label{appendix:prompt}
We used the prompt template reported in Figure~\ref{fig:prompt} for the privacy policy analysis with Gemini 3.1 Pro and GPT-5.4.
Each policy was provided as an HTML attachment.

\begin{figure}[!t] 
\centering
\footnotesize 
\fbox{
\begin{minipage}{0.95\columnwidth}
\raggedright %
You are a knowledgeable, helpful, and honest assistant.
You have deep expertise in current privacy regulations, including the GDPR, CCPA, and other U.S.\ state privacy laws.

\medskip
{\bf **Task**} 

I will provide an attachment of a therapy app's privacy policy.
Determine whether it includes information about: (1) Collecting camera, photo, image, or video data; (2) Collecting microphone, audio, voice, or recording data; (3) Specific named AI providers; (4) Specific named trackers; (5) Sharing data with advertisers.

\medskip
{\bf **Classification Rules**}

\textit{Named AI Providers:} Flag ``1'' only if the policy explicitly names a specific AI company or model (OpenAI, Anthropic, Claude, ChatGPT, GPT-4, Google Gemini, Groq, Llama, Mistral, Cohere).
Do NOT flag generic phrases (``artificial intelligence,'' ``AI-powered,'' ``machine learning'').
The app developer itself is NOT an AI provider.

\textit{Named Trackers:} Flag ``1'' only if the policy names specific tracking/analytics services (Google Analytics, Firebase, Mixpanel, AppsFlyer, Facebook Pixel, Amplitude, Segment, Hotjar, Braze, Adjust, Branch).
Do NOT count social login, cloud hosting, payment processors, or generic phrases (``cookies,'' ``web beacons,'' ``analytics services'') without company names.

\textit{Camera/Photo:} Flag ``1'' if the policy mentions collecting photographs, images, video, camera access, profile pictures, facial data, or biometric images.
Check CCPA disclosure tables for ``sensory data'' or ``audio/visual information.''

\textit{Microphone/Audio:} Flag ``1'' if the policy mentions collecting audio, voice recordings, speech, microphone access, or call recordings.
Voice-to-text transcription should still be flagged since audio is collected, even if temporarily.

\textit{Advertising:} Flag ``1'' if the policy mentions sharing data with advertisers, targeted/interest-based/ behavioral advertising, ad networks, or marketing partners.
Flag ``0'' only if they explicitly deny advertising AND there is no other indication.

\medskip
{\bf **Response Format**}

Camera/Photo: [flag, texts] \\
Microphone/Audio: [flag, texts] \\
Named AI Providers: [flag, texts] \\
Named Trackers: [flag, texts] \\
Advertising: [flag, texts]

Where flag = ``1'' (present) or ``0'' (absent), and texts = list of exact quotes or ``None.''
\end{minipage}
}
\caption{ LLM extraction prompt template for privacy policy analysis.}
\label{fig:prompt}
\end{figure}

\section{MobSF Composite Security Scores}\label{app:mobsf-scores}
\begin{figure}[!t] 
\centering
\includegraphics[width=0.99\columnwidth]{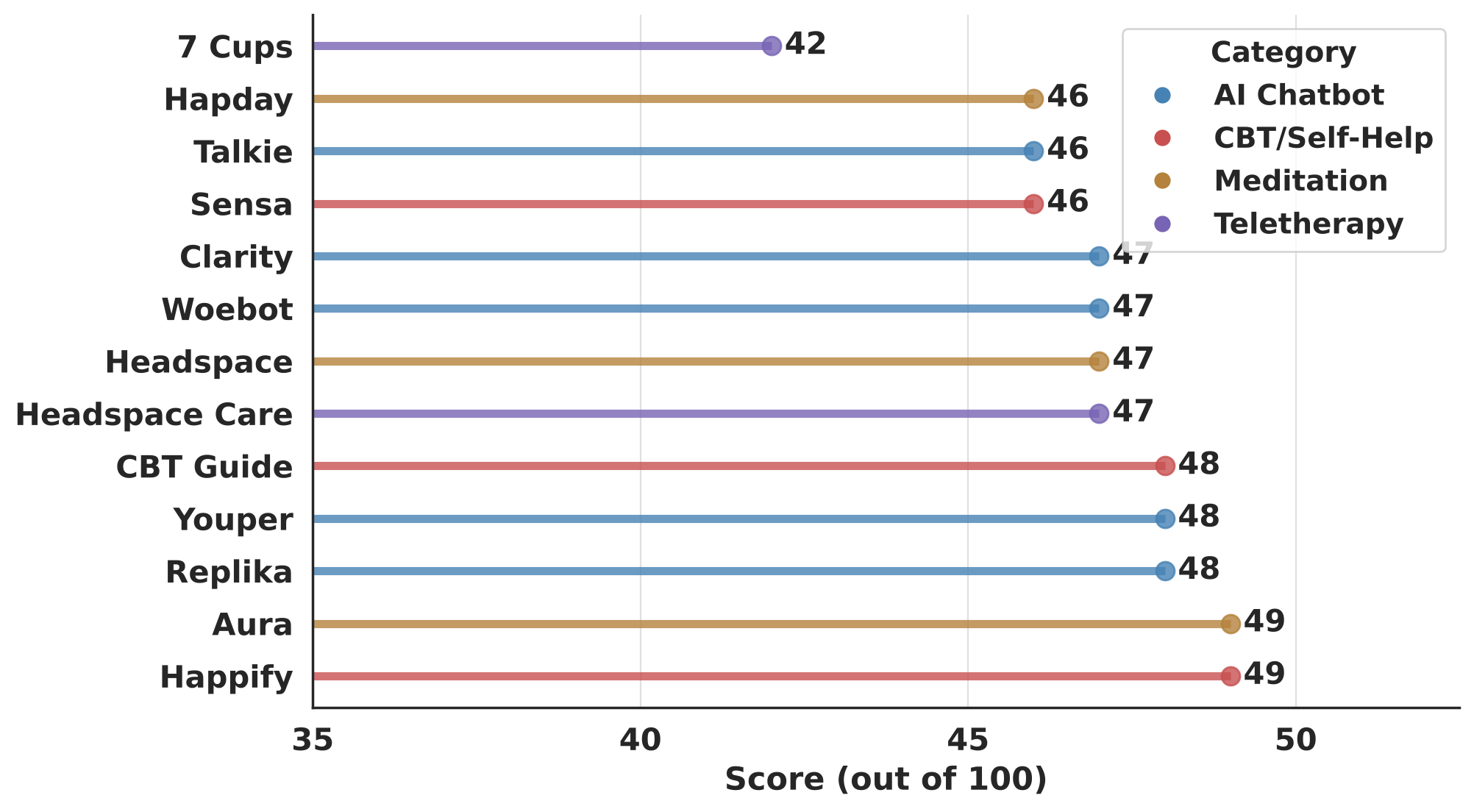}
\caption{Apps with MobSF composite scores below 50/100 (13 of 25, 52\%), ordered by score.}
\label{fig:security_low}
\end{figure}

MobSF computes a composite App Security Score in a [0-100] range as a weighted count of the high, warning, and secure labels it assigns, rather than from the per-rule Common Vulnerability Scoring System (CVSS) values it also records~\cite{mobsf-appsec-score}.
Higher scores indicate better security.
We report it here as descriptive context for ranking apps within our corpus, consistent with prior work using MobSF on mental-health and e-payment apps~\cite{iwaya2022privacy, kishnani2023assessing}, rather than as a validated security metric.
Scores range from $42$ (7~Cups) to $56$ (Rocky, Yana, Limbic), with $13$ apps ($52$\%) scoring below $50$, as reported in Figure~\ref{fig:security_low}.
The average is $49.9$ and the median $49$.
The spread is narrow (14 points), so the $50$-point threshold separates apps that differ little; this is one reason we treat the score as context rather than as a result.
The score also has blind spots.
As discussed in Section~\ref{sec:dyn-ai}, Rocky attains the joint-highest score in our corpus while shipping a permanent AWS access key inside the app.

\section{Data Deletion Contact Information}\label{app:contact}
Table~\ref{tab:deletion-contacts} lists deletion contact information for each app.

\begin{table}[!t] 
\centering
\setlength{\tabcolsep}{2pt}
\begin{tabular}{lll}
\toprule
\textbf{App} & \textbf{Method} & \textbf{Contact} \\
\midrule
7~Cups & Email & privacy@7Cups.com \\
Ash & In-app + Email & support@slingshotai.com \\
Aura & Email & hello@aurahealth.io \\
BetterUp & In-app + Web & privacyrequests.betterup.co \\
CBT Guide & None stated & -- \\
Clarity & Email & support@inquiryhealth.com \\
Hapday & Email & contact@haaaaaaappy.com \\
Happify & Email & support@mydario.com \\
Headspace & Email & help@headspace.com \\
Headspace Care  & Email & help@headspace.com \\
Limbic & Email & data.enquiries@limbic.ai \\
MindDoc & In-app + Email & support@minddoc.de \\
Neurofit & Email & support@neurofit.app \\
Noah & Email & sophia@heynoah.ai \\
Pi & In-app (type ``!delete'') & privacy@inflection.ai \\
Replika & In-app + Email & privacy@replika.com \\
Rocky & In-app + Email & hello@rocky.ai \\
Rosebud & Email & support@rosebud.app \\
Sensa & Email & hello@sensa.health \\
Stoic & Email & s@getstoic.com \\
Talkie & In-app + Email & feedback@subsup.com \\
Woebot & Email & privacy@woebothealth.com \\
Wysa & In-app + Email & hello@wysa.com \\
Yana & Email/Mail & contacto@yana.com.mx \\
Youper & Web + Email & youper.ai/data-erasure \\
\bottomrule
\end{tabular}
\caption{Data deletion mechanisms and contact information.}
\label{tab:deletion-contacts}
\end{table}

\end{document}